\documentclass[lettersize,journal]{IEEEtran}
\usepackage[colorinlistoftodos, textsize=small]{todonotes}
\usepackage{amsmath,amsfonts}
\usepackage{soul}
\usepackage{adjustbox}
\usepackage{tabularx}
\usepackage{algorithm,algorithmic}
\usepackage{array}
\usepackage{caption}
\usepackage[colorlinks,
            linkcolor=blue,       
            anchorcolor=blue,  
            citecolor=blue,        
            ]{hyperref}
\usepackage[caption=false,font=normalsize,labelfont=sf,textfont=sf]{subfig}
\usepackage{textcomp}
\usepackage{stfloats}
\usepackage{url}
\usepackage{verbatim}
\usepackage{graphicx}
\usepackage{cite}
\usepackage{booktabs} 
\usepackage{balance}
\usepackage{multirow}
\usepackage[normalem]{ulem}

\useunder{\uline}{\ul}{}
\def\BibTeX{{\rm B\kern-.05em{\sc i\kern-.025em b}\kern-.08em
    T\kern-.1667em\lower.7ex\hbox{E}\kern-.125emX}}
\begin{document}

\title{A Visual Perception-Based Tunable Framework and Evaluation Benchmark for H.265/HEVC ROI Encryption}

\author{Xiang Zhang, Geng Wu, Wenbin Huang, Daoyong Fu, Fei Peng, Zhangjie Fu

\thanks{This work was supported in part by the National Natural Science Foundation of China under Grant 62202234, 62401270; China Postdoctoral Science Foundation under Grant 2023M741778.

Xiang Zhang, Geng Wu, Wenbin Huang, Daoyong Fu, and Zhangjie Fu are with the Engineering Research Center of Digital Forensics, Ministry of Education, Nanjing University of Information Science and Technology, Nanjing, Jiangsu 210044, China (e-mail: zhangxiang@nuist.edu.cn; 202312490279@nuist.edu.cn; wenbinhuang@nuist.edu.cn; fudymo@hotmail.com; fzj@nuist.edu.cn).

Fei Peng is with the School of Artificial Intelligence, Guangzhou University, Guangzhou, Guangdong 510006, China (e-mail: eepengf@gmail.com).
}}

\markboth{Journal of \LaTeX\ Class Files,~Vol.~14, No.~8, August~2021}%
{Shell \MakeLowercase{\textit{et al.}}: A Sample Article Using IEEEtran.cls for IEEE Journals}


\maketitle

\begin{abstract}
ROI selective encryption, as an efficient privacy protection technique, encrypts only the key regions in the video, thereby ensuring security while minimizing the impact on coding efficiency. However, existing ROI-based video encryption methods suffer from insufficient flexibility and lack of a unified evaluation system. In particular, the lack of standard testing platforms is a major obstacle to the development of this field. To address these issues, we propose a visual perception-based tunable framework and evaluation benchmark for H.265/HEVC ROI encryption. Our scheme introduces three key contributions: 1) A ROI region recognition module based on visual perception network is proposed to accurately identify the ROI region in videos. 2) A three-level tunable encryption strategy is implemented while balancing security and real-time performance. 3) A unified ROI encryption evaluation benchmark is developed to provide a standardized quantitative platform for subsequent research. This triple strategy provides new solution and significant unified performance evaluation methods for ROI selective encryption field. Experimental results indicate that the proposed benchmark can comprehensively measure the performance of ROI selective encryption and provide a complete testing solution for this field. Compared to existing ROI encryption algorithms, our proposed enhanced and advanced level encryption exhibit superior performance in multiple performance metrics. In general, the proposed framework effectively meets the privacy protection requirements in H.265/HEVC and provides a reliable solution for secure and efficient processing of sensitive video content.
\end{abstract}


\section{Introduction}
\IEEEPARstart With the rapid development of video technologies, such as home health monitoring~\cite{khosravy2022model}, intelligent video surveillance~\cite{tian2021robust}, and remote conferencing, video data has become a primary medium for information exchange. Although these video streams provide great convenience and enable intelligent analysis, they often contain a substantial amount of sensitive information, particularly facial features, raising significant privacy concerns. Traditional encryption schemes, such as full-frame encryption, are effective privacy protection solutions. However, such methods completely disrupt the structure and content of the video, making it unsuitable for any subsequent analysis ~\cite{hirose2022anonymization}. Homomorphic encryption allows computation to be performed directly on encrypted data, but its extremely high computational complexity (e.g., exponential operations) makes it impractical for processing high-bitrate video streams in real-time systems~\cite{kim2023optimized}.

\begin{figure}[!t]
    \centering
    \includegraphics[width=0.7\columnwidth]{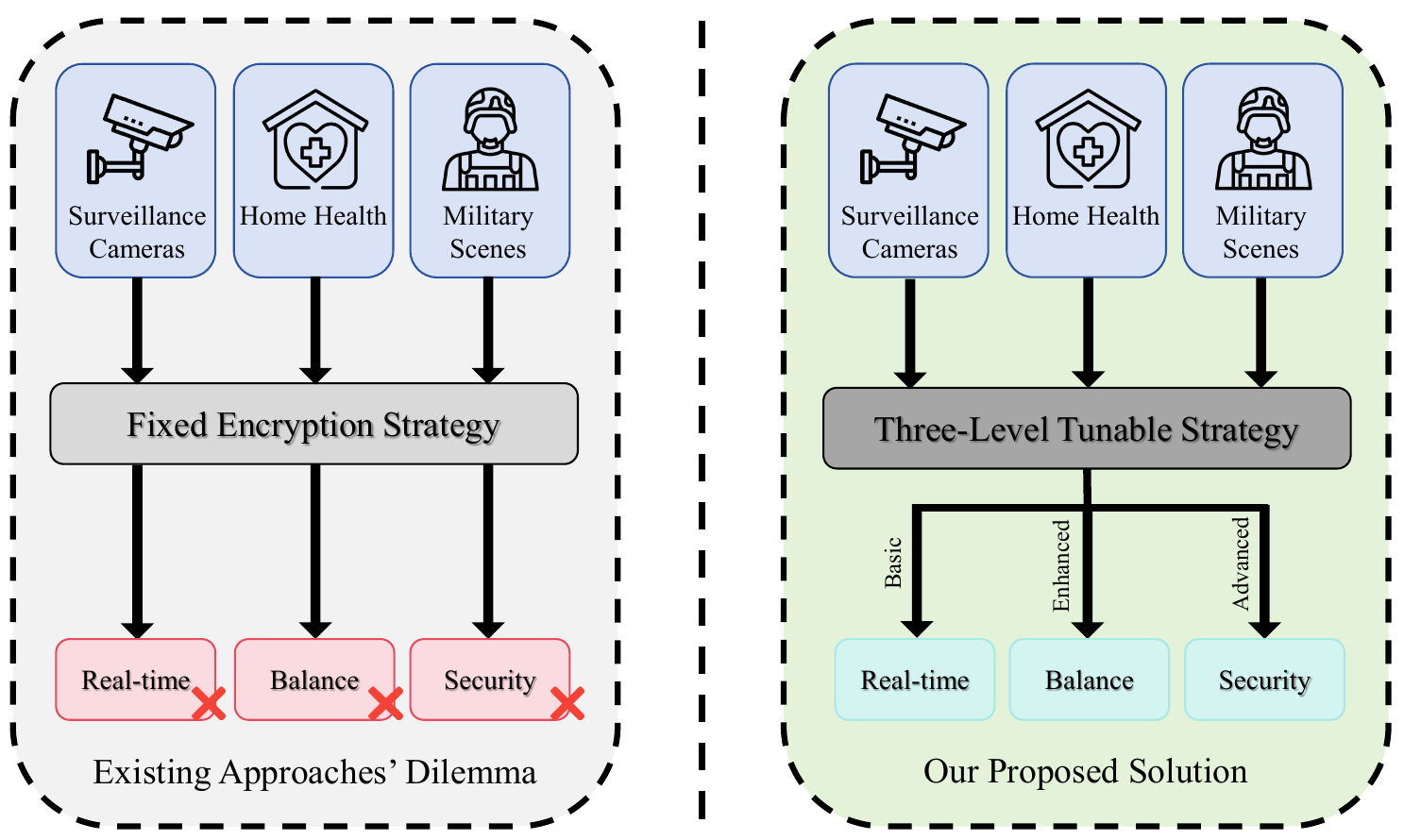}
    \caption{\small Comparison between fixed encryption strategy and proposed three-level encryption strategy}
    \label{fig11}
\end{figure}

Therefore, protecting sensitive information while maintaining the usability of video data for automated analysis tasks (such as behavior recognition) remains a significant challenge. To address this issue, researchers have proposed Region of Interest (ROI) encryption. The core concept of ROI encryption is to selectively encrypt only sensitive regions within video frames while keeping non-sensitive areas unencrypted. Early naive ROI encryption methods performed encryption directly on raw pixel data prior to compression, severely disrupting spatial correlations within the video~\cite{carrillo2008compression,dufaux2008h,dufaux2008scrambling,hosny2022privacy,li2024ppl}. As a result, the compression efficiency of subsequent codecs (e.g., H.265/HEVC) was significantly reduced, and bitstream compatibility could no longer be guaranteed~\cite{stutz2011survey}.

Consequently, Region of Interest (ROI) Selective Encryption (SE)~\cite{xu2013roi}, a strategy that integrates encryption with video encoding, has become a key research focus in the field of video security~\cite{sheng2024content}. Since H.265/HEVC is currently one of the most widely adopted video coding standards~\cite{van2013encryption}, ROI SE techniques based on H.265/HEVC have become a major research focus in the field of video encryption~\cite{peng2013roi,im2024cabac,yu2023coding,taha2018end,sohn2009privacy,farajallah2015roi}. Such approaches preserve bitstream format compliance with minimal bitrate overhead and enable standard decoders to reconstruct non-ROI regions without requiring access to the encryption key. Despite the great potential of ROI SE based on H.265/HEVC, existing algorithms still face two major challenges: \textbf{(1) Lack of Flexibility in Encryption Schemes:} Existing methods typically employ a fixed combination of syntax elements when encrypting ROI regions. For example, Peng \textit{et al}. \cite{peng2013roi} encrypts CAVLC, IPM, and MVD, achieving relatively high security but introducing approximately an 8\% increase in bitrate. In contrast, Sohn \textit{et al}. \cite{sohn2009privacy} only encrypts the transform quantization coefficients, resulting in a low bit rate increase. However, the limited distortion leads to insufficient overall security. These single encryption strategies are difficult to apply in complex real-world scenarios, since different application scenarios exhibit diverse requirements for security and real-time performance, which can be shown in Fig. \ref{fig11}. Therefore, existing methods fail to achieve a dynamic balance between security and real-time performance. \textbf{(2) Lack of a Unified and Standardized Evaluation Benchmark:} The absence of a unified evaluation benchmark remains a key challenge in the ROI encryption field. TABLE~\ref{metric_summary} summarizes several representative ROI encryption studies based on H.264/AVC and H.265/HEVC, including their datasets, experimental focus, and comparison algorithm. As shown in TABLE~\ref{metric_summary}, the selected datasets of these studies vary widely, ranging from 10 QCIF sequences with unspecified sources to standardized test sets such as Xiph.org, JCT-VC common test conditions, and SPEVI. Moreover, there is substantial diversity in the performance metrics, some studies primarily consider efficiency-related indicators, such as encoding time \cite{peng2013roi,yu2023coding}, while others focus on visual quality, bit rate change, and key security \cite{taha2018end,sohn2009privacy,farajallah2015roi}. In addition, we can also find from the table that the comparison algorithms are also inconsistent, with most studies lacking direct comparisons, except for a few that reference prior methods.

\begin{table}[!t]
\centering
\caption{\small Summary of some typical ROI SE schemes}
\label{metric_summary}
\renewcommand{\arraystretch}{1.3}
\resizebox{\linewidth}{!}{%
\Large
\begin{tabular}{l l p{4.5cm} p{5.5cm} p{5.5cm}}
\hline
\hline
\textbf{Scheme} & \textbf{Standard} & \textbf{Dataset} & \textbf{Experimental focus} & \textbf{Comparison algorithm} \\
\hline
Peng et al. \cite{peng2013roi} & H.264/AVC & 10 QCIF sequences (unspecified source) & Key security, Encoding time, Bit rate change & None \\
Yu et al. \cite{yu2023coding} & H.265/HEVC & Xiph.org video test media & PSNR, SSIM, IoU, Encoding time & None \\
Taha et al. \cite{taha2018end} & H.265/HEVC & JCT-VC common test conditions & PSNR, SSIM, Bit rate change, Key security & Van \textit{et al.} \cite{van2013encryption}, Boyadjis \textit{et al.} \cite{boyadjis2016extended} \\
Sohn et al. \cite{sohn2009privacy} & H.264/SVC & SPEVI dataset & Bit rate change & None \\
Farajallah et al. \cite{farajallah2015roi} & H.265/HEVC & JCT-VC common test conditions & Bit rate change & None \\
\hline
\hline
\end{tabular}%
}
\end{table}

This arbitrariness in metric selection, combined with the lack of transparency in testing methodologies, has led to the absence of a unified, comprehensive, and quantifiable benchmark in the field. Therefore, we propose a visual perception-based tunable framework and evaluation benchmark for H.265/HEVC ROI encryption. Firstly, a visual perception network is used to identify ROI coordinates and segment the video frame into ROI and non-ROI region. Then, a three-level tunable ROI encryption strategy (the basic, enhanced, and advanced level) is applied to encrypt ROI region of the video based on the user requirements. Finally, a unified ROI encryption evaluation benchmark is developed to better measure the performance of ROI encryption. We hope that the proposed benchmark can provide a reference testing platform for this field. In summary, the main contributions of this paper are as follows:

\begin{itemize}
\item \textbf{A ROI Region Recognition Module Based on Visual Perception Network.} We propose a ROI region recognition module effectively identify ROI region in video by training a visual perception network, significantly enhancing the accuracy of ROI recognition.

\item \textbf{Three-Level Tunable ROI Selective Encryption Strategy.} We present a three-level tunable ROI selective encryption strategy to balance security and real-time performance. Each level of encryption strategy perturbs different syntax elements. Users can flexibly select the encryption strength according to available device resources.

\item \textbf{A Unified ROI Encryption Evaluation Benchmark.} We pioneer a comprehensive evaluation benchmark which contains the evaluation of ROI region recognition accuracy and ROI region perturbation effect. It enables the quantitative comparison of different ROI selective encryption schemes and provides an important reference for future optimization efforts.

\item \textbf{Extensive Experimental Validation with Different Comparison Algorithms.} We conduct extensive experiments using the proposed ROI encryption evaluation benchmark and compare our method with Start-of-the-Art approaches. The experimental results demonstrate that the proposed scheme effectively satisfies diverse requirements in terms of security and real-time performance.
\end{itemize}

The remaining content is as follows: Section \ref{sec2} presents related work. Section \ref{sec3} describes the preliminaries. Section \ref{sec4} details the proposed video encryption scheme. Section \ref{sec5} provides the proposed ROI encryption evaluation benchmark. Section \ref{sec6} introduces experimental results and analysis. Finally, conclusions are drawn in Section \ref{sec7}.

\section{Related Work}\label{sec2}
\subsection{ROI Naive Encryption}
In ROI naive encryption, the ROI within the video is separately extracted and treated as the key region that needs to be protected. Traditional encryption algorithms, such as the Advanced Encryption Standard (AES), are usually applied to encrypt the data within this region. Carrillo \textit{et al}. \cite{carrillo2008compression} proposed protecting the ROI by encrypting the macroblock in the ROI before compression. However, this encryption process results in an approximately 23\% increase in video bit rate. Dufaux \textit{et al}. \cite{dufaux2008h} introduced a privacy protection method for video surveillance, utilizing a random permutation method to pseudo-randomly permute the alternating current (AC) coefficients within the ROI block. This algorithm also leads to a significant increase in bit rate.

To address the bit rate increment issue, Dufaux \textit{et al}. \cite{dufaux2008scrambling} proposed randomly selecting portions of the video bitstream for pseudo-random inversion, achieving a zero bit rate increase. However, this method compromises the compatibility of the video encoding format, leading to video decoding failure. Hosny \textit{et al}. \cite{hosny2022privacy} suggested a privacy protection approach for surveillance videos. They use YOLOv3 to detect the face region, then encrypting each pixel within the extracted ROI. This method does not fully obfuscate the original data, allowing attackers to potentially recover sensitive information. Li \textit{et al}. \cite{li2024ppl} proposed the PPL-enc scheme, which uses SOLOv2 to accurately identify privacy regions and combines pixel-level encryption with attribute-based encryption. This scheme also offers good real-time performance with minimal impact on storage and transmission.

Although ROI naive encryption can protect privacy by directly encrypting sensitive regions, it faces the technical challenges of a significant decrease in video coding efficiency, with bit rate increases ranging from 8\% to 23\% , Therefore, ROI selective encryption is proposed to solve the problems.

\subsection{ROI Selective Encryption}
Selective encryption takes into account both security and efficiency and has become a research focus in the field of video privacy protection. Peng \textit{et al}. \cite{peng2013roi } proposed a solution based on Flexible Macroblock Ordering (FMO) and chaotic encryption. FMO is first used to map the face region to a specific slice group, and then the chaotic encryption algorithm is used to encrypt the region. This method achieves a certain balance between security and encryption efficiency. However, the division of FMO slice groups may cause the encrypted region to exceed the ROI range. Im \textit{et al}. \cite{im2024cabac} combined Mask Region-Based Convolutional Neural Network (Mask R-CNN) object detection and Context-Adaptive Binary Arithmetic Coding (CABAC) to encrypt the detected privacy region, which not only ensures the precise positioning of the ROI, but also maintains the compatibility of the coding process. However, some ROI information may not be fully encrypted in low-bitrate videos.

To solve the above problems, Yu \textit{et al}. \cite{yu2023coding} proposed a coding unit (CU) level ROI encryption scheme based on H.265/HEVC. By using YOLOv4 \cite{bochkovskiy2020yolov4} to detect the face region and selectively encrypt the corresponding CU, it achieved relatively accurate ROI protection. However, it will cause the problem of reduced compression efficiency in high-resolution videos. Taha \textit{et al}. \cite{taha2018end} adopted a tile-level encryption strategy, using the independence of tiles in H.265/HEVC to encrypt only the key syntax elements in the tile containing the ROI, thereby achieving real-time protection. However, the limitations of tile division may cause the encrypted region to deviate from the actual ROI. Sohn \textit{et al}. \cite{sohn2009privacy} divided the face region into foreground and background based on scalable video coding technology, and adopted an encryption method of randomly flipping the transform coefficient sign for the foreground region. It has shortcomings in encryption propagation control. The AES-based scheme proposed by Farajallah \textit{et al}. \cite{farajallah2015roi} adopts full encryption and selective encryption strategies respectively. Selective encryption shows good format compatibility and encryption efficiency by encrypting only the key syntax elements in CABAC encoding, but it still has limitations in the precise control of the ROI region.

\subsection{ROI Region Recognition}\label{VPN}
Both ROI naive encryption and ROI selective encryption rely on the accurate identification and effective extraction of the ROI, making the precision of the ROI region recognition technology crucial. Early ROI encryption schemes predominantly relied on traditional ROI region recognition methods, such as face detection based on skin color models or using the OpenCV library \cite{viola2001rapid}. These methods identify ROI regions through color space conversion and feature matching, offering advantages in terms of low computational complexity. However, detection accuracy decreases significantly when encountering challenges such as lighting variations, object occlusion, or overlapping objects \cite{jiao2019survey}. Additionally, these methods have difficulty adapting to morphological changes in the object, which can lead to a mismatch between the encrypted region and the actual sensitive region.

In recent years, deep learning-based ROI region recognition technology has gradually become the dominant choice in ROI encryption due to its high-precision detection capabilities. He \textit{et al}. \cite{he2017mask} proposed Mask R-CNN. It is a two-stage ROI region recognition algorithm that combines object detection and instance segmentation. It excels in accurately capturing the contours of objects, making it particularly suitable for privacy regions with irregular shapes. However, its computational complexity is high, which may hinder the efficiency of real-time encryption. Wang \textit{et al}. \cite{wang2020solov2} proposed SOLOv2, which is an instance segmentation-based ROI region recognition model, achieves efficient object segmentation through dynamic convolution and mask prediction. It significantly reduces computational complexity while maintaining high accuracy, making it suitable for real-time ROI encryption in high-resolution videos. Redmon \textit{et al}. \cite{redmon2018yolov3} proposed YOLOv3, which is a single-stage ROI region recognition algorithm with high speed and real-time performance. By incorporating multi-scale prediction and feature pyramid network, YOLOv3 improves detection accuracy for small and dense objects. However, YOLOv3 is slightly less accurate than two-stage detection algorithms, which may cause the encrypted region to slightly exceed the actual sensitive ROI region.

\section{Preliminaries}\label{sec3}
\subsection{Object Detection Based on Visual Perception Network}\label{sec3a}
\begin{figure}[htbp]
    \centering
    \includegraphics[width=0.8\columnwidth]{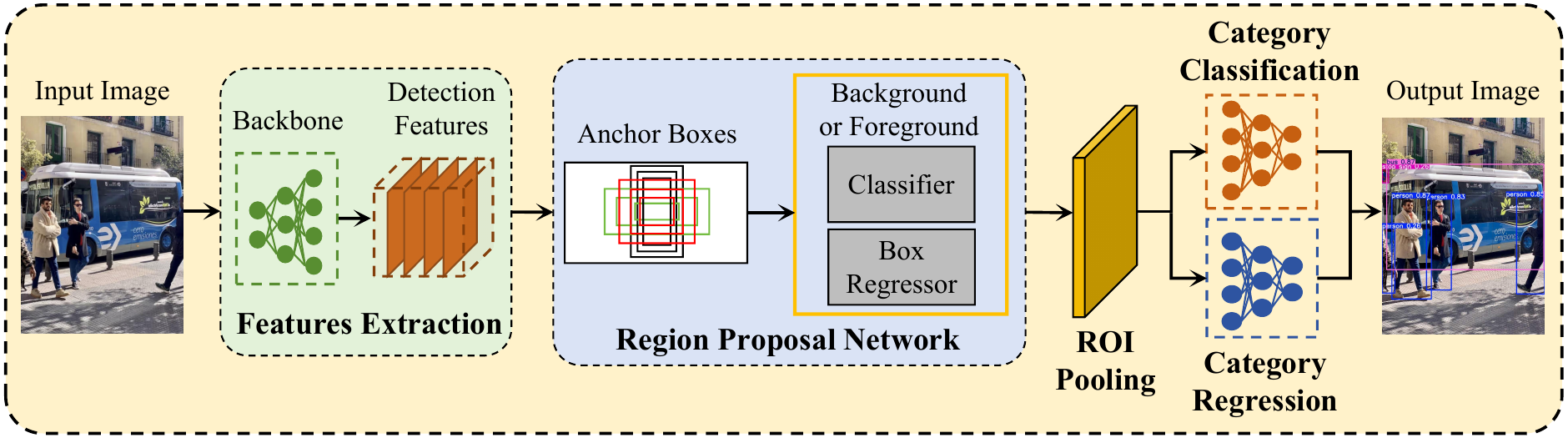}
    \caption{\small The architecture of visual perception network}
    \label{fig2}
\end{figure}

Object detection technology plays a crucial role in video content analysis and encryption systems. To maintain both high detection accuracy and real-time performance, we employ a visual perception network to accurately extract sensitive regions, providing reliable input for subsequent ROI encryption. The architecture of this network is illustrated in Fig. \ref{fig2}.

The visual perception network first utilizes backbone to extract detection features from the input image. Backbone consists of multiple convolutional and pooling layers, enabling the network to capture both local and global features, thus generating high-dimensional feature maps\cite{krizhevsky2012imagenet}. The detection features are then input into Region Proposal Network (RPN) \cite{ren2015faster} to generate multiple anchor boxes through a sliding window mechanism. These anchor boxes vary in scale and aspect ratio, enabling the network to accommodate objects of different sizes and shapes. Afterwards, the multiple anchor boxes are input into a foreground-background classifier and box regressor. Among them, the foreground-background classifier is used to determine whether the anchor box is a foreground (object) or a background and its loss function is:
\begin{equation}
L_{\text{cls}}^{\text{RPN}}=-\frac{1}{N}\sum_{i=1}^N\left[p_i\log\hat{p}_i+(1-p_i)\log(1-\hat{p}_i)\right],
\end{equation}
where \( N \) is the number of anchor boxes, \( p_i \) is the true label of anchor box \( i \) (1 for foreground, 0 for background), and \( \hat{p}_i \) is the probability predicted by RPN that anchor box \( i \) corresponds to the foreground. while the foreground-background box regressor is used to fine-tune the position of the anchor box in order to generate more accurate candidate regions and its loss function is:
\begin{equation}
L_{\text{reg}}^{\text{RPN}}=\frac{1}{N}\sum_{i=1}^N\mathrm{smooth}_{L1}(t_i-\hat{t}_i)],
\end{equation}
where \( t_i \) is the true offset of anchor box \( i \). \( \hat{t}_i\) is the offset  of anchor box \( i \) predicted by RPN, and $\mathrm{smooth}_{L1}(\cdot)$ is the smooth L1 loss function.
Through the above process, RPN generates a series of high-quality candidate regions. Then, the ROI pooling layer maps the candidate regions into a fixed-size classification vector. Finally, the classification vector is entered into category classification network and category regression network. Among them, category classification network classifies the classification vector to determine which category it belongs to and its loss function is:
\begin{equation}
L_{\mathrm{cls}}=-\frac{1}{M}\sum_{j=1}^M\log\hat{p}_{j,c_j},
\end{equation}
where \( M \) is the number of ROIs, \( c_j \) is the true category of ROI \( j \), and \( \hat{p}_{j,c_j}\) is the probability that ROI \( j \) belongs to category \( c_j \). While category regression network is used to fine-tune the position of the classified ROI to generate a more accurate bounding box. The category regression loss is:
\begin{equation}
L_{\mathrm{reg}}=\frac{1}{M}\sum_{j=1}^M\mathrm{smooth}_{L1}(t_j-\hat{t}_j),
\end{equation}
where \( t_j \) is the actual offset of ROI \( j \), and \( \hat{t}_j \) is the offset predicted by the category regression network.

Through the features extraction, region proposal network, ROI pooling, category classification and regression, the visual perception network accurately extracts sensitive regions from the image, enabling reliable input for the tunable ROI encryption process.

\subsection{Analysis of Syntax Elements in H.265/HEVC}\label{3B}
The syntax elements is the basic units used to represent video information during the encoding process. In H.265/HEVC, the video frame is compressed into a bitstream after prediction, transformation, quantization and entropy coding. In this process, a large number of syntax elements are obtained. TABLE \ref{tab1} provides the main syntax elements in H.265/HEVC. These syntax elements are binarized using three methods: $k$-order Exponential Golomb ($\mathrm{EG}_k$) coding, $k$-order Truncated Rice ($\mathrm{TR}_k$) coding, and Fixed-Length (FL) coding\cite{flynn2015overview}. After the binarization stage, CABAC is applied. H.265/HEVC has two entropy coding modes in CABAC: regular mode and bypass mode. In the regular mode, the encoding of each binary value is based on an adaptively updated probability model. In the bypass mode, the encoding of syntax elements is based on fixed probabilities.

\begin{table}[htbp]
\centering
\caption{\small Main syntax elements of H.265/HEVC}
\begin{tabular}{cccc}
\toprule
\textbf{Syntax element} & \textbf{Entropy mode} & \textbf{Binarization} & \textbf{Category} \\ \midrule
Luma IPM       & Regular        & $\mathrm{TR}_k$          & \multirow{7}{*}{Prediction} \\
Chroma IPM     & Regular        & $\mathrm{TR}_k$          &                             \\
Merge index    & Regular, Bypass & FL           &                             \\
MVD sign       & Bypass         & FL           &                             \\
MVD value      & Bypass         & $\mathrm{EG}_k$          &                             \\
MVPIdx         & Regular        & FL           &                             \\
RefFrmIdx      & Regular, Bypass & $\mathrm{EG}_k$          &                             \\ \cmidrule{1-4}
Residual sign  & Bypass         & FL           & \multirow{3}{*}{Residual}   \\
Residual value & Bypass         & $\mathrm{TR}_k$          &                             \\
Delta QP value & Regular, Bypass & $\mathrm{EG}_k$          &                             \\ \cmidrule{1-4}
SAO parameter  & Bypass         & $\mathrm{EG}_k$          & Filtering                   \\ \bottomrule
\end{tabular}
\label{tab1}
\end{table}

While encrypting the syntax elements in regular mode will inevitably result in a change in the bit rate, and encrypting syntax elements in bypass mode will not increase the bit rate \cite{liu2010survey}. Based on this, we analyze the impact of encryption of each syntax element. As shown in TABLE \ref{tab1}, syntax elements are divided into three categories: prediction, residual, and filtering. In video coding, decoded pixels are primarily restored by prediction and residual data. Thus, encrypting prediction and residual syntax elements will cause a strong perturbation effect\cite{stutz2011survey}. Furthermore, according to the analysis in \cite{tang2022format}, SAO parameters do not significantly contribute to visual distortion. To enhance efficiency, it is recommended not to encrypt these parameters. In conclusion, encrypting IPM, MVD, Residual, and Delta QP offers a higher scrambling effect, and encrypting IPM, Merge index, MVPIdx, RefFrmIdx, and Delta QP inevitably change the bit rate. Additionally, the encryption must strictly adhere to the encoding rules of syntax elements to ensure that the encrypted bitstream remains compatible with the H.265/HEVC format.

\section{Proposed Three-level Tunable ROI Selective Encryption Scheme for H.265/HEVC}\label{sec4}
\begin{figure}[htbp]
    \centering
    \includegraphics[width=0.8\columnwidth]{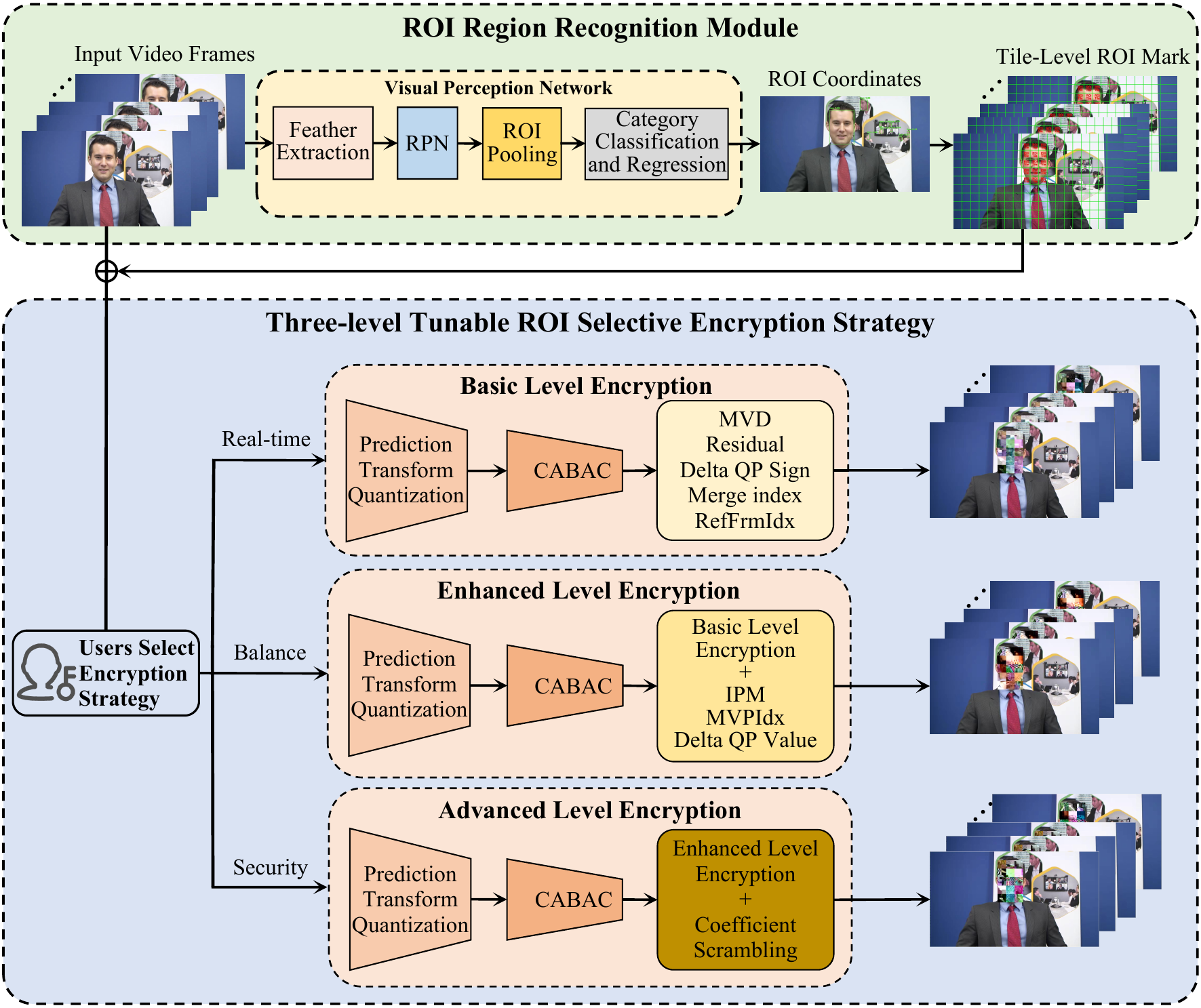}
    \caption{\small The framework of the proposed encryption strategy}
    \label{fig3}
\end{figure}

Based on the characteristics of syntax elements, we propose a graded encryption strategy for H.265/HEVC with three levels of encryption, including basic level, enhanced level, and advanced level. This strategy allows for the flexible selection of different encryption methods based on specific needs. The complete framework of the proposed three-level encryption strategy is shown in Fig. \ref{fig3}. In the ROI region recognition module, a visual perception network first extracts the ROI coordinates from each input video frame. These coordinates are then converted into tile-level ROI masks. Finally, based on specific application requirements, the proposed three-level tunable ROI selective encryption strategy is subsequently applied to encrypt the syntax elements within the marked tile-level ROI regions. Therefore, our encryption framework consists of two core components: ROI region recognition module and three-level tunable ROI selective encryption. 

\subsection{ROI Region Recognition Module}\label{sec4a}

In the ROI recognition process, we treat ``face" as the target and the recognition process is as follows:
\begin{equation}
    \text{ROI\_Coords} = {VPN}(F_n),
\end{equation}
where $F_n$ represents the current input frame, \(VPN(\cdot)\) represents a pre-trained object detection model based on visual perception network which is described in Section \ref{sec3a}. $\text{ROI\_Coords}$ is the output of the model, which is the ROI coordinate file.

Upon detection, $\text{ROI\_Coords}$ is stored as text coordinates in the format \(\text{frameID}:[i, (x_1^i, y_1^i, x_2^i, y_2^i)]\), where \(i\) denotes the frame index, and \((x_1^i, y_1^i)\) and \((x_2^i, y_2^i)\) represent the pixel coordinates of the top-left and bottom-right corners of the ROI region in the $i^{th}$ frame, respectively. Then, we classify Tiles in each frame into ROI Tiles and non-ROI Tiles based on $\text{ROI\_Coords}$. Afterwards, the encoder configuration file is modified to enable Tile partition, configured in a fixed-size mode with a uniform Tile size of \(32 \times 32\). The classification of Tiles is expressed as:
\begin{equation}
\text{curTile} =
\begin{cases}
\text{ROI-Tile}, & if\,\,\,\, mark(\text{curTile}) = 1 \\
\text{nonROI-Tile}, & \text{otherwise},
\end{cases}
\end{equation}
where \(mark(\cdot)\) function evaluates whether the current Tile's coordinate range overlaps with the $\text{ROI\_Coords}$, returning a Boolean value according to the following formula:
\begin{equation}
mark(\text{curTile}) =
\begin{cases}
1, & \begin{aligned}
& if\,\,\,\, (x^i_{\text{1\_tile}}, y^i_{\text{1\_tile}}, x^i_{\text{2\_tile}}, y^i_{\text{2\_tile}}) \\
& \cap (x_1^i, y_1^i, x_2^i, y_2^i) \neq \emptyset
\end{aligned} \\
0, & \text{otherwise},
\end{cases}
\end{equation}
where \((x^i_{\text{1\_tile}}, y^i_{\text{1\_tile}}, x^i_{\text{2\_tile}}, y^i_{\text{2\_tile}})\) denotes the boundary coordinates of the current Tile in the $i^{th}$ frame, and \(\cap\) is the intersection sign. After classifying all tiles, we proceed with encrypting the video frames. If the currently Tile is an \text{ROI-Tile}, the proposed three-level ROI selective encryption strategy is applied to the CUs within the Tile. Otherwise, the current Tile is encoded normally.

\subsection{Three-level Tunable ROI Selective Encryption}
In our strategy, a binary key generation method based on Advanced Stream Encryption-Counter Mode (ASE-CTR) \cite{lipmaa2000ctr} is first employed to generate the binary key stream $S$:
\begin{equation}
S = \mathrm{ASE}(K, \mathrm{CTR}),
\label{eq:encryption}
\end{equation}
where \( \mathrm{ASE}(K, \mathrm{CTR}) \) represents the pseudo-random binary key stream generated by encrypting with \( K \) and counter \( \mathrm{CTR} \), the counter \( \mathrm{CTR} \) is incremented after encrypting each syntax element to ensure that the key stream for each syntax element is unique. After generating the binary key stream sequence $S=\{s_1,s_2,...,s_{kl}\}$, where $kl$ is the length of \( S\), it is used to encrypt the binarized syntax elements in each CU of ROI-Tile, which is identified in Section \ref{sec4a}. The following section provides a detailed introduction to the three-level tunable ROI selective encryption strategy.

\subsubsection{Basic Level Encryption}
The basic level encryption aims to ensure real-time performance and zero bit increment, while providing basic visual distortion effects. It is suitable for scenarios with high real-time requirements. This strategy involves encrypting certain syntax elements such as MVD sign and value, residual sign and value, the suffix of Delta QP, Merge index, and the suffix of RefFrmIdx in bypass mode.

\noindent\textbf{Encryption of MVD Sign and Value.}
MVD represents the difference between the motion vector of the current prediction unit and its predicted value, comprising horizontal and vertical components, each consisting of a sign and a value. These components are encoded separately. The signs are binarized using FL coding and encrypted as:
\begin{gather}
    encMVDHsign = MVDHsign \oplus s_i,\,\,\,\,\! 0 \leq s_i \leq 1, \\
    encMVDVsign = MVDVsign \oplus s_i,\,\,\,\,\! 0 \leq s_i \leq 1,
\end{gather}
where \( MVDHsign \) and \( MVDVsign \) are the signs of the horizontal and vertical components of current MVD, respectively. \( s_i \) is encryption key extracted from \( S \). While MVD value is binarized using $\mathrm{EG}_1$ and its suffix is encrypted as:
\begin{equation}encMVDsuffix=MVDsuffix\oplus s_i,\,\,\,\,\! 0 \leq s_i \leq 1.\end{equation}

\noindent\textbf{Encryption of Residual Sign and Value.}
The residual structure is similar to MVD, consisting of a sign and a value, which are also encoded separately. The sign \(CoefSign\) of each non-zero coefficient is encoded independently into the bitstream. Here, \(CoefSign=0\) indicates a positive number, and \(CoefSign=1\) indicates a negative number. \(CoefSign\) is binarized using FL coding and encrypted as:
\begin{equation}
    \mathit{encCoefSign} = \mathit{CoefSign} \oplus s_i,\,\,\,\,\! 0 \leq s_i \leq 1.
\end{equation}

The residual value \( absCoefLevel\) consists of two parts: \( CoefbaseLevel\) and \( CoefremainingLevel\). 
\( CoefbaseLevel \) is not encrypted since it is encoded by regular mode. While \( CoefremainingLevel\) is binarized using $\mathrm{TR}_k$ and encoded by bypass mode. As a result, its suffix $Coefsuffix$ is encrypted as:
\begin{equation}
    \mathit{encCoefsuffix} = \mathit{Coefsuffix} \oplus s_i,\,\,\,\,\! 0 \leq s_i \leq 1.
\end{equation}

\noindent\textbf{Encryption of Delta QP sign.}
The quantization parameter (QP) controls the compression quality and encoding bit rate. It is composed of a sign and a value. To maintain a consistent bit rate, only the sign $uiSign$ is selected for encryption. If \(uiSign =0\), the Delta QP is positive. If \(uiSign =1\), the Delta QP is negative. $uiSign$ is encrypted as:
\begin{equation}
\mathit{encuiSign} = \mathit{uiSign} \oplus \mathit{s_i},\,\,\,\,\! 0 \leq s_i \leq 1.
\end{equation}
 
\noindent\textbf{Encryption of Merge index.}
The merge mode is used for inter prediction. In inter prediction, a candidate motion vector reference list of length 5 is constructed directly from adjacent prediction units. The index number within this list is referred to as the Merge index, and its value ranges from 0 to 4. The Merge index is encoded using FL coding and encrypted as:
\begin{gather}
    \mathit{encUnaryIdx} = (\mathit{UnaryIdx} + s_i) \,\% \,5,\,\,\,\,\! 0 \leq s_i \leq 3.
\end{gather}

\noindent\textbf{Encryption of RefFrmIdx.}
In H.265/HEVC, the current frame predicts its content by reference frames. The encoder use reference frame index to represents each reference frame. We encrypt the suffix of reference frame index since it is binarized using $\mathrm{EG}_k$ and encoded by bypass mode. However, the upper limit of the reference frame index varies depending on the buffer. Therefore, the encryption of the reference frame index suffix $RFIS$ according to the current number of the frames in the buffer $RN$ with the maximal value 4 is:
\begin{equation}
\mathit{encRFIS} =
\begin{cases}
\mathit{RFIS} \oplus s_i,  &0 \leq s_i \leq 1, \,\,\,if\,\,\, RN = 2 \\
(\mathit{RFIS} + s_i) \,\% \,3, &0 \leq s_i \leq 1,\,\,\,if\,\,\, RN = 3 \\
\mathit{RFIS} \oplus s_i,  &0 \leq s_i \leq 3,\,\, \,if\,\,\, RN = 4.
\end{cases}
\end{equation}

\subsubsection{Enhanced Level Encryption}
The enhanced level encryption preserves the real-time advantages of basic level encryption while strengthening content protection by broadening the scope of encryption. Building on the basic level encryption, the enhanced level encryption includes the encryption of IPM, MVPIdx, and Delta QP value, which can produce significant visual disruption effects. Since the entropy coding mode for these elements is regular mode, inevitably resulting in an increase in the bit rate. The specific encryption methods for these syntax elements are described below.

\noindent\textbf{Encryption of Luma IPM.}
Luma IPM is a critical syntax element in H.265/HEVC. The total number of luma IPMs is 35. During the encoding process, a list contains three Most Probable Modes (MPMs) is first constructed. If the current luma IPM is in the list, a MPM index $MPMIdx$ is recorded. $MPMIdx$ is encoded by bypass mode and encrypted as:
\begin{equation}
\mathit{encMPMIdx} = (\mathit{MPMIdx} + s_i) \,\%\, 3, \quad 0 \leq \mathit{s_i} \leq 3.
\end{equation}

If the current luma IPM is not included in the MPMs list, the remaining 32 modes will be conveniently represented by a 5-bit codeword $LumaIPM$ and then encrypted as:
\begin{equation}
\mathit{encLumaIPM} = \mathit{LumaIPM} \oplus s_i, \quad 0 \leq \mathit{s_i} \leq 31.
\end{equation}

\noindent\textbf{Encryption of Chroma IPM.}
Chroma IPM has five prediction modes, and is recorded using the chroma IPM index $ChromaIPM$. The encryption of $ChromaIPM$ can significantly Perturb the video. Therefore, we encrypt it as:
\begin{equation}
\mathit{encChromaIPM} = (\mathit{ChromaIPM} + s_i) \,\%\, 3,  \,\,\,\,\,0 \leq \mathit{s_i} \leq 3.
\end{equation}

\noindent\textbf{Encryption of MVPIdx.}
MVP index is the syntax element in the Advanced Motion Vector Prediction (AMVP). When the AMVP mode is used, the motion vector is predicted from neighbor blocks, and a list of MVP candidates with a length of 2 is formed. The MVP index $MVPIdx$, that identifies the optimal MVP from this list, is binarized by regular mode and encrypted as:
\begin{equation}
\mathit{encMVPIdx} = \mathit{MVPIdx} \oplus s_i, \quad 0 \leq s_i \leq 1.
\end{equation}

\noindent\textbf{Encryption of Delta QP Value.}
To avoid decoding failure, encryption is Delta QP Value \( \mathit{DQP} \) requires ensuring that the encrypted value does not exceed maximum Delta QP value \( \mathit{MaxDQP} \). The encryption process for \( \mathit{DQP} \) is defined as:
\begin{equation}
\begin{aligned}
&\mathit{encDQP} = \\ 
&(\mathit{DQP} \oplus s_i) \,\% \,\mathit{MaxDQP} + 1, \quad0 \leq s_i \leq MaxDQP-1.
\end{aligned}
\end{equation}

\subsubsection{Advanced Level Encryption}
The edge contains the highest concentration of video information, making its protection essential \cite{zhang2013edge}. Existing ROI encryption typically perturbs syntax elements but neglects edge information. To address this, we proposed an improved coefficient scrambling algorithm based on \cite{peng2019tunable} for the edge region in the advanced level encryption, building on the enhanced level encryption. The brief steps of the coefficient scrambling algorithm are as follows:

\textit{Step} 1: The Canny operator $Canny(\cdot)$ is applied to filter the luma component of a frame $FL$ to obtain the pixel edge information $FL^{\text{edge}}(x,y)$:
\begin{equation}
FL^{\text{edge}}(x,y)=Canny(FL(x,y)).
\end{equation}

\textit{Step} 2: The pixel edge information $FL^{\text{edge}}$ is binarized using the binarization function \(Bin(\cdot)\), resulting in the luma edge binary image $FL^{\text{bin}}$:
\begin{equation}
FL^{\text{bin}}(x,y) = Bin(FL^{\text{edge}}(x,y)).
\end{equation}

\textit{Step} 3: The TUs in \( FL_n^{\text{bin}} \) are classified into two categories based on the pixels \(  FL^{\text{bin}}_{\mathrm{TU}}(x,y) \) in the TUs:
\begin{equation}
\mathrm{TU} =
\begin{cases}
\mathrm{edgeTU,} & \displaystyle if\,\,\,\, \sum FL^{\text{bin}}_{\mathrm{TU}}(x,y) \neq 0 \\
\text{non-edgeTU,} & \text{otherwise}.
\end{cases}
\end{equation}

\textit{Step} 4: The $N_{nz}$ non-zero coefficients in an $\mathrm{edgeTU}$ are divided into two sets: the first $N_{nz}-1$ non-zero coefficients in scan order form the permuted set $C_{\text{permute}}$, and the last non-zero scanned coefficient is reserved as $LastCoef$ for reversible embedding in Step 5. To enhance security beyond the general-purpose generator in \cite{peng2019tunable}, a key-driven chaotic permutation is applied to $C_{\text{permute}}$. The procedure is as follows:
\begin{enumerate}
    \item Let $N_p = N_{nz}-1$ be the number of coefficients to permute. If $N_p \le 1$, skip this step.
    \item Use the secret key $K_c$ to generate the initial value $x_0$ and control parameter $r$ for a Logistic chaotic map: 
    \begin{equation}
    x_{i+1} = r \cdot x_i (1 - x_i).
    \end{equation}
    \item Iterate the chaotic map $N_p$ times to obtain a floating-point sequence $S = \{x_1, \dots, x_{N_p}\}$.
    \item Generate a permutation map $\pi = \{\pi(1), \dots, \pi(N_p)\}$ by sorting $S$ and recording the indices.
    \item Scramble the coefficients in $C_{\text{permute}} = \{c_1, \dots, c_{N_p}\}$ according to $\pi$ to obtain the scrambled set $C'_{\text{permute}} = \{c'_1, \dots, c'_{N_p}\}$:
    \begin{equation}
    c'_{\pi(k)} = c_k, \quad k = 1, \dots, N_p.
    \end{equation}
\end{enumerate}

\textit{Step} 5: Reversibly embed the sign into the last non-zero scanned coefficient \(LastCoef\) in the current TU:
\begin{equation}
LastCoef' =
\begin{cases} 
2 \times LastCoef - w, \quad if \,\,\,\, LastCoef > 0 \\
2 \times LastCoef + w, \quad if \,\,\,\, LastCoef < 0, 
\end{cases} \quad
\end{equation}
where \(w\) is a value used to record the TU type. If current TU is \(\mathrm{edgeTU}\), set \(w = 1\). Otherwise, set \(w = 0\).

\section{ROI Encryption Evaluation Benchmark}\label{sec5}
ROI encryption have recently garnered significant attention in the field of video encryption. However, the current research on ROI encryption often lacks comprehensive experimental evaluation, and the dataset, comparison algorithm, and evaluation methods used are not standardized. For instance,  Peng\cite{peng2013roi} assessed encoding time and compression ratio with no comparison algorithm in 10 QCIF sequences with unspecified sources. Yu\cite{yu2023coding} focused on testing IoU between the encrypted region and the ground truth, as well as encryption time, but did not include any comparisons on the Xiph.org dataset. Taha\cite{taha2018end} evaluated their scheme by testing bit rate change, PSNR and SSIM values with Van \textit{et al.} \cite{van2013encryption} and Boyadjis \textit{et al.} \cite{boyadjis2016extended} in the JCT-VC common test dataset. These evaluation methods are relatively simplistic, lacking sufficient persuasive power to demonstrate the superiority of the scheme. Furthermore, the inconsistency in dataset, comparison algorithm, and evaluation methods complicates effective horizontal comparisons and comprehensive evaluations between different schemes.

In conclusion, this field is in urgent need of a standardized evaluation benchmark. Therefore, we pioneer a new comprehensive \textbf{ROI encryption evaluation benchmark} to provide a standardized evaluation platform for different ROI encryption algorithms. We believe that the performance comparison of an ROI encryption algorithm depends on three key factors: \textit{datasets, comparison algorithm, and evaluation methods.} (All the contents of our proposed benchmark are available at \url{https://github.com/playboiwg/ROI-encryption-evaluation-benchmark/tree/main})

\subsection{Datasets}\label{dataset}
Existing ROI encryption studies exhibit considerable arbitrariness in dataset selection, often employing QCIF sequences of unclear origin or outdated formats, alongside diverse test sets from sources such as Xiph.org, JCT-VC, and SPEVI. This lack of unified standards severely impedes fair, comprehensive, and quantifiable cross-comparisons within the field. To address this issue, a comprehensive evaluation benchmark should first establish a standardized set of video datasets, ensuring that different encryption schemes can be objectively assessed and compared under identical testing conditions. Although there are many publicly available video datasets, not all of them contain ROI regions. Therefore, we compile and collect totally 55 YUV sequences that contain ROI regions from widely recognized public datasets commonly used in video coding and video processing, which can serve as standard test sets for future studies. TABLE~\ref{tab:video_datasets} summarizes these public datasets and their main characteristics. In the future, we will also further collect and expand our ROI encryption dataset.

\begin{table}[htbp]
\centering
\caption{\small Standard video datasets constructed for evaluation}
\label{tab:video_datasets}
\begin{tabularx}{\columnwidth}{l c c >{\raggedright\arraybackslash}X}
\hline
\hline
Dataset & Resolution & Quantity & Link \\
\hline
UVG Dataset &
4K &
7 & %
\url{https://ultravideo.fi/dataset.html} \\
Xiph.org Media &
QCIF -- 4K &
18 & %
\url{https://media.xiph.org/video/derf/} \\
MSU CVQAD &
240P -- 4K &
30 & %
\url{https://videoprocessing.ai/datasets/cvqad.html} \\
\hline
\hline
\end{tabularx}
\end{table}

\noindent\textbf{UVG Dataset.} We selected 7 sequences containing ROIs from the UVG dataset. These sequences were originally captured at 120~fps with a resolution of $3840\times2160$. They contain clear human subjects and specific motion patterns, such as \textit{Beauty} (close-up face) and \textit{Jockey} (high-speed human motion). These videos possess extremely high texture details, enabling the simulation of privacy protection requirements in ultra-clear surveillance or film production scenarios.

\noindent\textbf{Xiph.org Media.} We constructed a test set of 18 sequences containing ROIs from the Xiph.org repository, with resolutions ranging from $176\times144$ to $3840\times2160$ and frame rates ranging from 25 to 60~fps. Focusing on typical application scenarios such as video conferencing, news broadcasting, and traffic surveillance, this selection covers diverse ROIs ranging from single faces (e.g., \textit{Foreman}, \textit{Akiyo}) to multi-target pedestrians and vehicles (e.g., \textit{Bus}, \textit{Coastguard}).

\noindent\textbf{MSU CVQAD.} We incorporated 30 video clips containing ROIs from the MSU CVQAD database, with resolutions ranging from $480\times240$ to $3840\times2160$ and a frame rate of 30~fps. We specifically selected videos featuring high-density crowds and complex street backgrounds. These scenes are often accompanied by illumination changes and object occlusions. Testing on these sequences allows evaluation of the algorithm’s encryption performance in complex environments, simulating real urban security surveillance scenarios.

\subsection{Comparison Algorithms}\label{Algorithms}
To enable a comprehensive and unified evaluation, our benchmark incorporates six representative ROI selective encryption schemes reported in the existing literature, covering multiple coding standards such as H.264/AVC, H.265/HEVC, and H.266/VVC. All representative algorithms are summarized and analyzed to establish a more complete comparison framework. The implementation details and typical advantages and limitations of each method are presented in TABLE~\ref{tab:roi_methods}, and their brief introductions are as follows:
\begin{table*}[!htbp]
\centering
\caption{\small Summary of representative ROI SE schemes}
\label{tab:roi_methods}
\resizebox{\textwidth}{!}{
\begin{tabular}{p{2.2cm} p{2.2cm} p{5.2cm} p{3.2cm} p{3.2cm}}
\hline
\hline
Scheme & Standard & Implementation detail & Advantage & Limitation \\
\hline

Peng \textit{et al}. \cite{peng2013roi}. & H.264/AVC & Uses FMO to separate ROI and non-ROI, encrypts IPM and MVD symbol bits using a chaotic system & High format compatibility and strong security & High bitrate increase \\

Kim \textit{et al}. \cite{kim2007selective} & H.264/SVC & Uses PRNG-generated sequence to encrypt MVD within ROI & High real-time performance & Limited security \\

Yu \textit{et al}. \cite{yu2023coding} & H.265/HEVC & Encrypts IPM, MV symbol bits, and TC symbol bits using AES-CFB generated keys & Fine-grained encryption, high IoU accuracy & High bitrate increase \\

Taha \textit{et al}. \cite{taha2018end} & H.265/HEVC & Encrypts IPM, MVD, MV symbol bits, and TC using a chaotic system & Good real-time performance & High bitrate increase \\

Im \textit{et al}. \cite{im2024cabac} & H.266/VVC & Encrypts MVD and residual signals using AES-128 & Good real-time performance & Possible error propagation \\

Im \textit{et al}. \cite{im2023roi} & H.266/VVC & Encrypts MVD, residual signals, and QP using AES-128 & Higher security & Reduced real-time performance \\

\hline
\hline
\end{tabular}
}
\end{table*}

\noindent\textbf{Peng \textit{et al}. \cite{peng2013roi}}
This comparative algorithm is designed for H.264/AVC, it employs FMO to isolate ROI macroblocks into independent slice groups and employs a chaotic system to encrypt the IPM and MVD.

\noindent\textbf{Kim \textit{et al}. \cite{kim2007selective}}
This comparative algorithm is designed for H.264/SVC, it uses a sequence generated by a pseudo-eandom number generator to selectively encrypt the MVD within the ROI layers.

\noindent\textbf{Yu \textit{et al}. \cite{yu2023coding}}
This comparative algorithm is designed for H.265/HEVC, it utilizes the AES-CFB mode to generate keystreams for the joint encryption of IPM, MV sign bits, and TC sign bits.

\noindent\textbf{Taha \textit{et al}. \cite{taha2018end}}
This comparative algorithm is based on the H.265/HEVC standard, which employs a chaotic system to encrypt IPM, MVD, MV sign bits, and TC within the ROI.

\noindent\textbf{Im \textit{et al}. \cite{im2024cabac}}
This comparative algorithm is designed for the H.266/VVC standard, and it uses the AES-128 algorithm within the CABAC entropy coding process to encrypt MVD and residual signals in the ROI.

\noindent\textbf{Im \textit{et al}. \cite{im2023roi}}
This algorithm serves as an extension of \cite{im2024cabac}, further utilizing the AES-128 algorithm to encrypt MVD, residual signals, and QP to enhance encryption strength.

To ensure fairness and standardization in comparisons, we recommend selecting the corresponding algorithms from the aforementioned representative schemes as comparison baselines according to the adopted video coding standard.

\subsection{Evaluation Methods}\label{sec5.3}
The IoU between the encrypted regions and the detected ROIs can quantify how accurately sensitive areas are covered by the encryption. Meanwhile, the perturbation effect within the extracted ROI regions, which affects whether the content in these regions is sufficiently perturbed to become unrecognizable. They are the two parts of evaluation methods in our proposed benchmark: the evaluation of ROI encryption fineness and ROI region perturbation effect.

\subsubsection{The Evaluation of ROI Encryption Fineness}\label{IOU}
In this section, we evaluate the fineness of ROI encryption at the pixel level. Specifically, the actually encrypted regions of encoded video and the ROI region ground truth are represented as pixel sets, denoted as $E$ and $G$, respectively, where $E$ represents the set of encrypted pixels and $G$ represents the set of ROI region ground truth pixels. The encryption fineness is measured using the IoU between these two sets, defined as:
\begin{equation}
\text{IoU} = \frac{|E \cap G|}{|E \cup G|},
\end{equation}
where $|E \cap G|$ denotes the number of pixels in the intersection of the encrypted and ground truth, and $|E \cup G|$ denotes the number of pixels in their union. The closer the IoU value is to 1, the higher the overlap between the encrypted region and the ground truth, indicating that the encryption covers the sensitive areas more precisely and achieves higher fineness. Conversely, a lower IoU value indicates incomplete or excessive encryption, reflecting insufficient fineness.

To provide a comprehensive assessment for an entire video sequence, the IoU can be computed for each frame and averaged over all frames to obtain the average encryption fineness, denoted as $\text{IoU}_{\text{avg}}$:
\begin{equation}
\text{IoU}_{\text{avg}} = \frac{1}{N} \sum_{i=1}^{N} \frac{|E_i \cap G_i|}{|E_i \cup G_i|},
\end{equation}
where $N$ is the total number of frames in a video, and $E_i$ and $G_i$ represent the encrypted pixel set and ground truth pixel set in the $i$-th frame, respectively. This metric quantifies the pixel-level coverage precision of the encrypted regions and provides a unified and intuitive standard for evaluating the fineness of ROI encryption algorithms.

\subsubsection{The Evaluation of ROI Region Perturbation Effect}
The evaluation of ROI region perturbation effect includes two parts: the analysis of subjective vision and the analysis of objective indicators.
\paragraph{Analysis of Subjective Vision}\label{SV}
To assess the perceptual impact of encryption on the ROI region, we conduct subjective vision comparison by presenting encrypted video frames to the observers. This leverages participant feedback to evaluate the level of perturbation in the ROI region. Specifically, at least two video sequences with different resolutions and texture complexities are selected. For each video sequence, we choose the first and last frames, and show their original frames, encrypted frames (the proposed algorithm), and the corresponding decrypted frames (the proposed algorithm), which can refer to Fig. \ref{fig5}. Subsequently, the whole encrypted frames, encrypted images of ROI region, and encrypted edge images of ROI region are compared with those produced by SOTA comparative algorithms, which can refer to Fig. \ref{fig7}, Fig. \ref{fig8}, and Fig. \ref{fig9}. Furthermore, various subjective and intuitive visualization examples can be included in the future to illustrate the encryption effectiveness of the proposed algorithm.

\paragraph{Analysis of Objective Indicators}
Unlike global encryption algorithms, evaluating an ROI encryption algorithm presents unique challenges, as it encrypts only the ROI region of a video frame, and the extracted ROI region may vary across different algorithms. Thus, the objective indicators used for global encryption cannot be directly applied. To address this issue, we designed an accurate ROI region extraction criterion. This criterion enables precise extracting of all ROI regions based on different ROI region recognition algorithms. Then, we apply various objective evaluation indicators to these regions to assess the security of the encryption algorithm. 

\noindent\textbf{ROI Region Extraction Criterion.}
In video coding standards, videos are typically compressed at the level of coding units rather than individual pixels. Therefore, to accurately assess the perturbation effect within the ROI region, evaluation metrics should be applied to the actual coding units that have been encrypted. To this end, we propose an ROI region extraction criterion based on basic coding units, enabling precise identification of all encrypted units and ensuring that the evaluation results correspond strictly to the actual encrypted regions.

\textit{Definition of Basic Coding Unit:}
We first define $U$ as a ``basic coding unit" in video coding. The basic coding unit is a general concept, it represents the basic block structure of video encoding. Meanwhile, its specific form varies across different video coding standards. For instance, it corresponds to a Macroblock (MB) in H.264/AVC and to a Coding Unit (CU) in H.265/HEVC.

\textit{Identification of Target Unit Set:}
For any video frame, let $S_{\text{all}}$ denote the set of all basic coding units in the frame:
\begin{equation}
S_{\text{all}} = \{ U_1, U_2, \dots, U_n \},
\end{equation}
where $n$ is the total number of basic coding units in the frame. Given the ROI coordinates $ROI\_Coords$ output by the ROI recognition module, the encoder first traverses $S_{\text{all}}$, if the current basic coding unit $U_j$ overlaps with the ROI region, it is marked as a ROI basic coding unit. All marked ROI basic coding units form the ROI unit set $S_{\text{ROI}}$:
\begin{equation}
S_{\text{ROI}} = \{ U_j \in S_{\text{all}} \mid \text{Area}(U_j) \cap \text{Area}(ROI\_Coords) \neq \emptyset \}.
\end{equation}
where $\text{Area}(\cdot)$ denotes a function that returns pixel coordinate sets. The intersection operator $\cap$ returns the same pixel coordinates between these pixel coordinate sets, and $\emptyset$ represents an empty set.

\textit{Data Extraction:}
To compute objective evaluation metrics, two types of data are extracted according to $S_{\text{ROI}}$:
\begin{itemize}
    \item $D_{\text{ori}}(U_j)$: Y, U, V components of $U_j$ of the original video in $S_{\text{ROI}}$.
    \item $D_{\text{enc}}(U_j)$: Y, U, V components of $U_j$ of the encrypted video in $S_{\text{ROI}}$.
\end{itemize}

These data are aggregated to form the original ROI pixel set $P_{\text{ori}}$ and the encrypted ROI pixel set $P_{\text{enc}}$:
\begin{equation}
P_{\text{ori}} = \bigcup_{U_j \in S_{\text{ROI}}} D_{\text{ori}}(U_j), \quad
P_{\text{enc}} = \bigcup_{U_j \in S_{\text{ROI}}} D_{\text{enc}}(U_j).
\end{equation}
where the symbol $\bigcup$ denotes the union of sets, which combines the Y, U, and V components into a complete ROI pixel set.

Finally, all objective evaluation metrics are computed by comparing $P_{\text{ori}}$ and $P_{\text{enc}}$. By adopting the above ROI region extraction criterion, the data used for assessment strictly correspond to the actual encrypted ROI regions. Therefore, the evaluation results are significantly more accurate and reliable.

\noindent\textbf{Objective Evaluation Indicators.}
After extracting the ROI region of each frame based on the proposed ROI region extraction criterion, we can integrate a comprehensive set of objective evaluation indicators to systematically assess the perturbation effect of ROI region. These indicators include: Peak Signal-to-Noise Ratio (PSNR), Structural Similarity Index (SSIM), Edge Difference Ratio (EDR), information entropy, Pixel Change Rate (NPCR), Unified Average Change Intensity (UACI), and Bit Rate Change. To provide a clear overview, TABLE \ref{tab3} summarizes the key objective indicators along with their calculation formulas and evaluation dimensions. The following provides a detailed interpretation of the indicators.

\begin{table*}[htbp]
\caption{\small Video encryption objective evaluation indicators}\label{tab3}
\resizebox{\linewidth}{!}{ 
\begin{tabular}{cccc}
\hline
\hline
Indicator & Brief description & Calculation formula & Evaluation dimension \\ \hline
\multirow{2}{*}{PSNR} & \multirow{2}{*}{\begin{tabular}[c]{@{}c@{}}Measure pixel-level noise of\\ original and encrypted videos.\end{tabular}} & \multirow{2}{*}{PSNR $= 20 \log_{10} \left( \frac{255}{\mathrm{MSE}} \right)$} & \multirow{2}{*}{Video Quality} \\
 &  &  &  \\
\multirow{2}{*}{SSIM} & \multirow{2}{*}{\begin{tabular}[c]{@{}c@{}}Evaluate structural similarity, \\ taking into account brightness and contrast.\end{tabular}} & \multirow{2}{*}{SSIM $= \frac{(2\mu_1\mu_2+c_{1})(2\sigma_1\sigma_2+c_{2})}{(\mu_1^{2}+\mu_2^{2}+c_{1})(\sigma_1^{2}+\sigma_2^{2}+c_{2})}$} & \multirow{2}{*}{Video Quality} \\
 &  &  &  \\
\multirow{2}{*}{EDR} & \multirow{2}{*}{\begin{tabular}[c]{@{}c@{}} Quantify edge distortion\\ introduced by encryption.\end{tabular}} & \multirow{2}{*}{EDR $= \frac{\sum_{i=1}^{h}\sum_{j=1}^{w}|PE(i,j)-CE(i,j)|}{\sum_{i=1}^{h}\sum_{j=1}^{w}|PE(i,j)+CE(i,j)|}$} & \multirow{2}{*}{Video Quality} \\
 &  &  &  \\
\multirow{2}{*}{\begin{tabular}[c]{@{}c@{}}Information\\ Entropy\end{tabular}} & \multirow{2}{*}{\begin{tabular}[c]{@{}c@{}}Evaluating the randomness \\ of encrypted data.\end{tabular}} & \multirow{2}{*}{H$(I) = -\sum_{j=1}^{2^L} p(I_j) \log_2 p(I_j)$} & \multirow{2}{*}{Security Strength} \\
 &  &  &  \\
\multirow{2}{*}{NPCR} & \multirow{2}{*}{Calculate the percentage of pixel change.} & \multirow{2}{*}{NPCR $= \left( \frac{\sum_{i=1}^h \sum_{j=1}^w D(i,j)}{h \times w} \right) \times 100\%$} & \multirow{2}{*}{Security Strength} \\
 &  &  &  \\
UACI & \begin{tabular}[c]{@{}c@{}}Measures the average \\ intensity of pixel differences.\end{tabular} & UACI $= \left( \frac{\sum_{i=1}^h \sum_{j=1}^w |C_1(i,j) - C_2(i,j)|}{255 \times h \times w} \right) \times 100\%$ & Security Strength \\
Bit rate change & \begin{tabular}[c]{@{}c@{}}Analyze the change in \\ bit rate after encryption.\end{tabular} & $\triangle$Bitrate$(\%)$ = $\frac{\text{Bitrate}_{\text{enc}} - \text{Bitrate}_{\text{ori}}}{\text{Bitrate}_{\text{ori}}} \times 100\%$ & Compression Efficiency \\ \hline
\hline
\end{tabular}%
}
\end{table*}

PSNR is a commonly used video quality evaluation metric. It quantifies the degree of video distortion by calculating the difference between the original video and the encrypted video. The lower the PSNR, the higher the security of the encrypted video. In the formula, Mean Squared Error (MSE) is the average pixel difference between the original and encrypted video frame, calculated as:\begin{equation}
\begin{aligned}
\mathrm{MSE} = \frac{1}{N} \sum_{i=1}^{N} (I_{\text{ori}}(i,j) - I_{\text{enc}}(i,j))^2,
\end{aligned}
\end{equation}
where $I_{\text{ori}}(i,j)$ and $I_{\text{enc}}(i,j)$ are the pixel values at position $(i,j)$ for the original and encrypted video frame, respectively, and $N$ is the total number of pixels in the video frame.

SSIM is a metric used to measure the similarity between video frames. It comprehensively evaluates the similarity of video frames from three aspects: brightness, contrast, and structure. The smaller the value of SSIM, the greater the difference between two video frames. In the formula, $\mu_1$ and $\mu_2$ are the average luminance of the original video frame and the encrypted video frame, $\sigma_1^2$ and $\sigma_2^2$ are the variances of the original video frame and the encrypted video frame, used to measure the contrast of the video frames. $\sigma_1\sigma_2$ is the covariance between the original video frame and the encrypted video frame. $c_1$ and $c_2$ are the constants. 

EDR is an important indicator to evaluate edge distortion of video encryption algorithms. The edge information of the encrypted video frame should be severely distorted so that it cannot be recognized. The higher the EDR value, the greater the edge distortion. In the formula, \( PE(i,j) \) and \( CE(i,j) \) are the edge pixel value of the original video frame and the encrypted video frame at position \( (i,j) \), respectively, and \( h \) and \( w \) are the height and width of the video frame.

Information entropy is used to measure the randomness and uncertainty of encrypted video. The closer the information entropy value is to the ideal value of 8, the stronger the randomness and uncertainty of the encrypted video. In the formula, \( L \) is the binary representation length of the pixel value, \( p(I_j) \) is the probability of the pixel value \( I_j \) appearing.

NPCR and UACI are two commonly used indicators for evaluating the ability of encryption schemes to resist differential attacks. The expected value of NPCR and UACI are 99.6094\% and 33.4635\%, respectively. In the formula of NPCR, \(D(i, j)\) is the difference matrix, defined as:
\begin{equation}
D(i,j) = 
\begin{cases}
0, & \text{if } C_1(i,j) = C_2(i,j) \\
1, & \text{if } C_1(i,j) \neq C_2(i,j),
\end{cases}
\end{equation}
where \(C_1(i,j)\) and \(C_2(i,j)\) represent the encrypted pixel value of two encrypted frame at position $(i,j)$.

Bit rate change is an important indicator to measure the encryption algorithm. Ideally, encryption will not cause an increase in bit rate. In the formula, Bitrate$_{\text{ori}}$ and Bitrate$_{\text{enc}}$ are the bit rates before and after encryption.

In summary, the seven objective indicators in the proposed benchmark measure the performance in terms of video quality, security strength, and compression efficiency, helping researchers effectively identify the advantages and disadvantages of various ROI encryption schemes on a standardized platform. In the future, more objective indicators can be added to enrich the proposed benchmark.

\section{Experimental Results and Analysis}\label{sec6}
\subsection{Experimental Setup}
The proposed ROI selective encryption scheme is implemented in the H.265/HEVC reference software HM 16.9. The experimental computer configuration is as follows: Intel (R) Core (TM) i7-6700 HQ, 2.6 GHz, 16 GB memory, with Windows 10 operating system, Microsoft Visual Studio 2010, MATLAB 2021a and OpenCV 2.4.7 installed. The visual perception network used in our ROI region recognition module is YOLOv8 \cite{wang2024dph} in the experiment. 

\begin{figure}[htbp]
    \centering
    \includegraphics[width=0.52\linewidth]{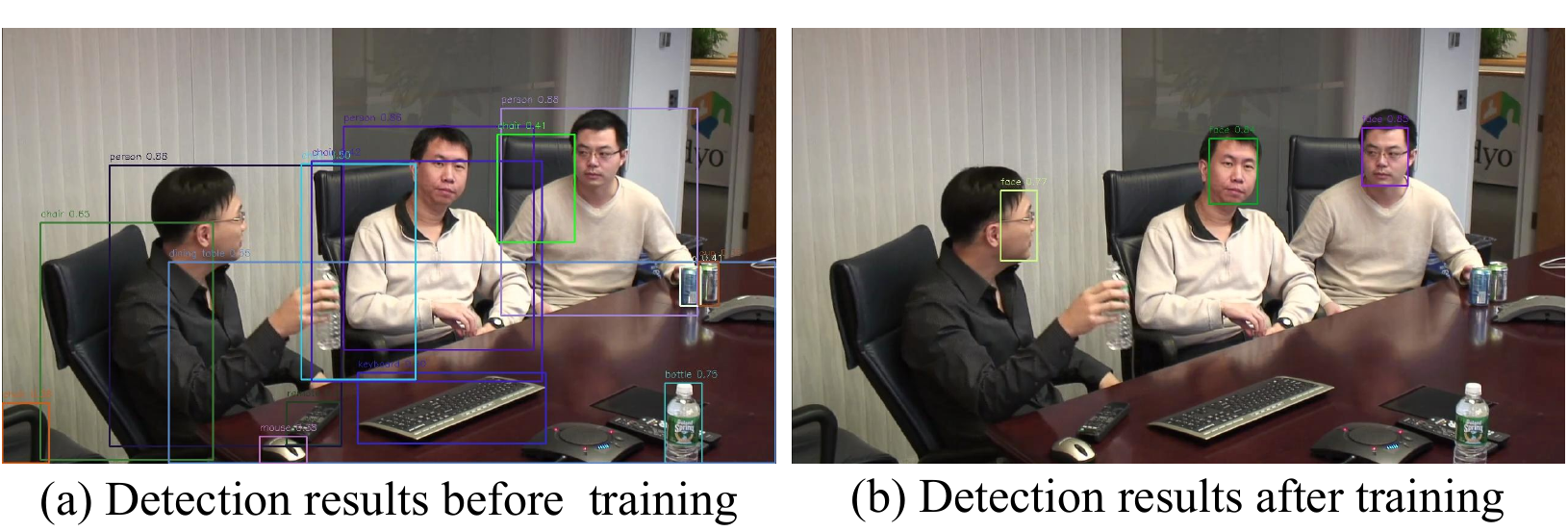}
    \caption{\small Detection results before and after training}
    \label{fig4}
\end{figure}

To enhance the sensitivity to the face region, the WiderFace dataset is used for targeted training. The dataset contains 32,203 images and 393,703 annotated faces, covering complex scenes such as scale, posture, and occlusion. After training, the output bounding box is strictly focused on the face region. The comparison of detection before and after training is shown in Fig. \ref{fig4}, which clearly shows the significant improvement of the module in face region detection after targeted training, providing high-precision ROI input for subsequent encryption.

The video sequences selected for the experiment are taken from the Xiph.org Media in TABLE~\ref{tab:video_datasets} of the proposed benchmark. Five sequences containing facial information were chosen for testing. The selected YUV video sequences are listed in TABLE \ref{tab4}, with resolutions ranging from $352\times288$ to $2560\times1600$, a frame rate of 60 fps, using the ``encoder\_lowdelay\_main'' profile, and a GOP structure of ``IBBB''. The QP is set to 8, 24, and 40, respectively. As shown in TABLE \ref{tab:roi_methods}, our benchmark provides six comparison algorithms. Since our algorithm is a ROI SE encryption algorithm for H.265/HEVC, we choose the two H.265/HEVC ROI encryption algorithms Yu \cite{yu2023coding} and Taha \cite{taha2018end} from TABLE \ref{tab:roi_methods} for comparison. We encrypt 50 frames of each sequence by the three schemes for comparison.
\begin{table}[htbp]
\centering
\caption{\small YUV test sequence}\label{tab4}
\begin{tabular}{ccccc}  
\hline
\hline
CLASS    & Resolution & Sequence   & FPS & Frames      \\ \hline
CLASS\_A & $352\times288$    & Akiyo       & 60  & 50   \\
CLASS\_B & $832\times480$    & PartyScene     & 60  & 50 \\
CLASS\_C & $1280\times720$   & Johnny        & 60  & 50  \\
CLASS\_D & $1980\times1080$  & Kimono        & 60  & 50  \\
CLASS\_E & $2560\times1600$  & PeopleOnStreet  & 60  & 50\\ \hline
\hline
\end{tabular}%
\end{table}

\subsection{Examples}
\begin{figure}[htbp]
    \centering
    \includegraphics[width=0.7\linewidth]{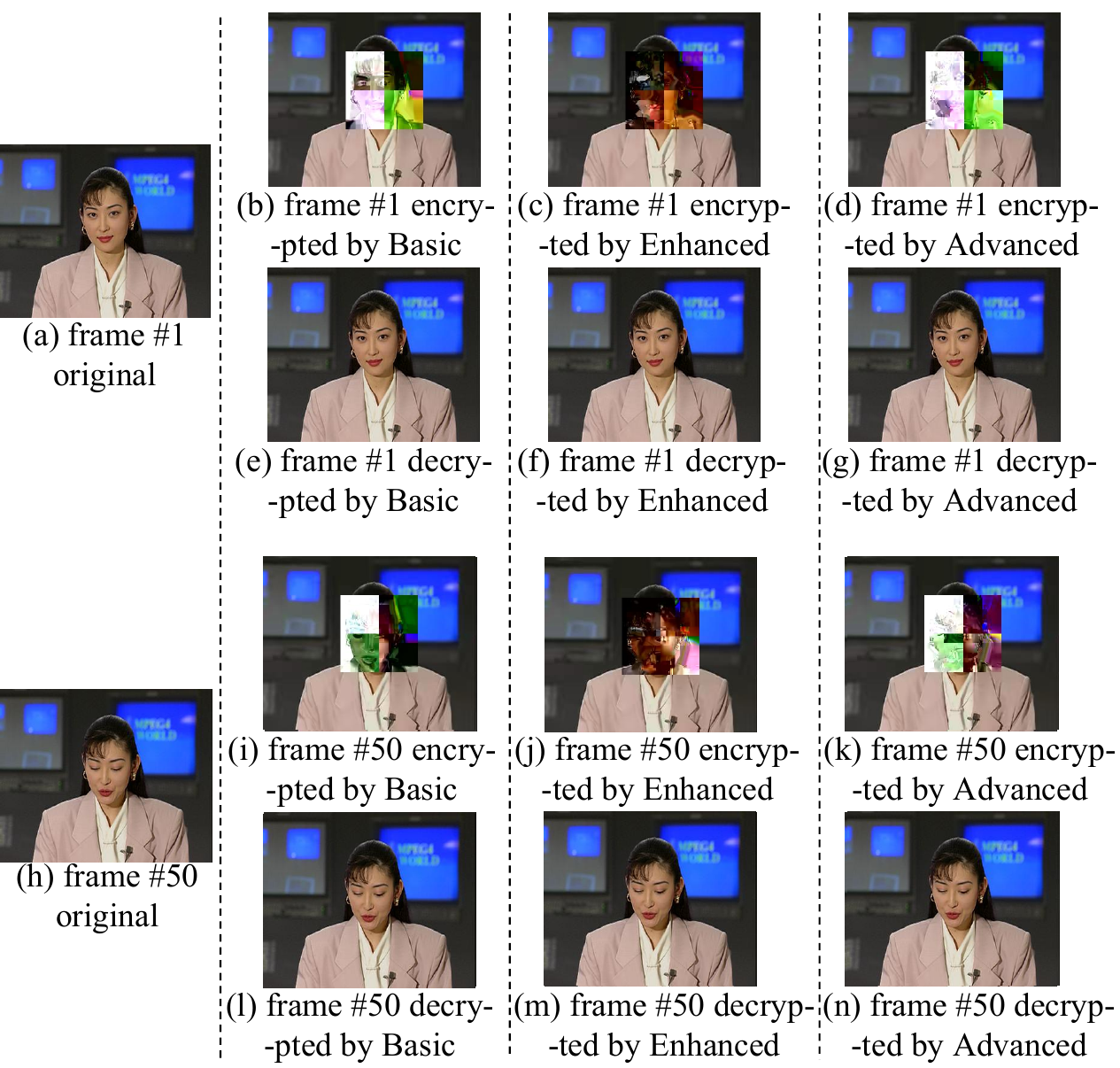}
    \caption{\small Some examples of proposed strategy in Akiyo sequence}
    \label{fig5}
\end{figure}

According to \ref{SV}, some sample frames of the proposed three-level encryption strategy in the Akiyo video sequence are shown in Fig. \ref{fig5}. 

As described in our proposed benchmark, the frame \#1 and \#50 of the video sequence are selected to intuitively demonstrate the encryption and decryption effects. The experimental results show that our encryption scheme performs effectively in H.265/HEVC codec. The encrypted frames show severe visual distortion in ROI region, and we can successfully recover the original Akiyo video sequence from the encrypted bitstream.

\subsection{Performance Analysis}
To evaluate the effectiveness of the proposed scheme, this section presents a comprehensive performance analysis conducted based on the proposed benchmark in Section \ref{sec5}. The evaluation is structured into two components: ROI encryption fineness and ROI region perturbation effect.
\subsubsection{Comparison of ROI Encryption Fineness}\label{detection model}
In this section, we evaluate the fineness of the proposed encryption framework and compare it with only Taha\cite{taha2018end}, since Yu \cite{yu2023coding} does not provide its ROI region recognition module. Meanwhile, we use our ROI region recognition module to detect Yu \cite{yu2023coding} for subsequent comparison of the perturbation effect. The evaluation is based on IoU$_{\text{avg}}$ defined in Section \ref{IOU}. A higher IoU$_{\text{avg}}$ indicates that the encryption scheme can more accurately map the detected ROI into the coding structure, thereby minimizing unnecessary encryption of non-target background pixels.
\begin{figure}[htbp]
    \centering
    \includegraphics[width=0.6\linewidth]{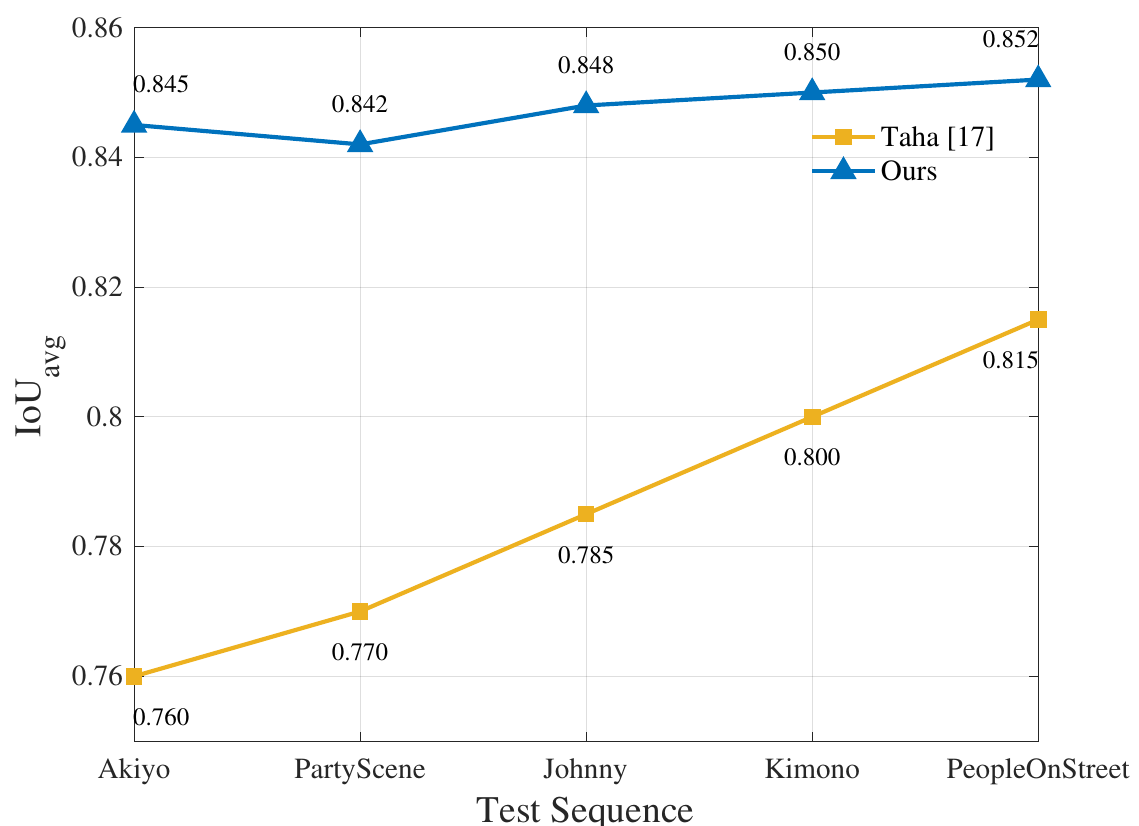}
    \caption{\small ROI encryption fineness results}
    \label{fig1}
\end{figure}

Fig. \ref{fig1} presents the IoU$_{\text{avg}}$ comparison results across all test sequences. As shown in the figure, our scheme employs Tile-level encryption with a fixed size of $16 \times 16$, achieving finer granularity than Taha's scheme, which uses fixed $32\times32$ Tiles and thus inevitably encrypts more non-ROI pixels.

\subsubsection{Comparison of ROI Region Perturbation Effect}
The comparison of ROI region perturbation effect includes two parts: comparison of subjective vision and comparison of objective indicators.

\begin{figure}[htbp]
    \centering
    \includegraphics[width=0.85\linewidth]{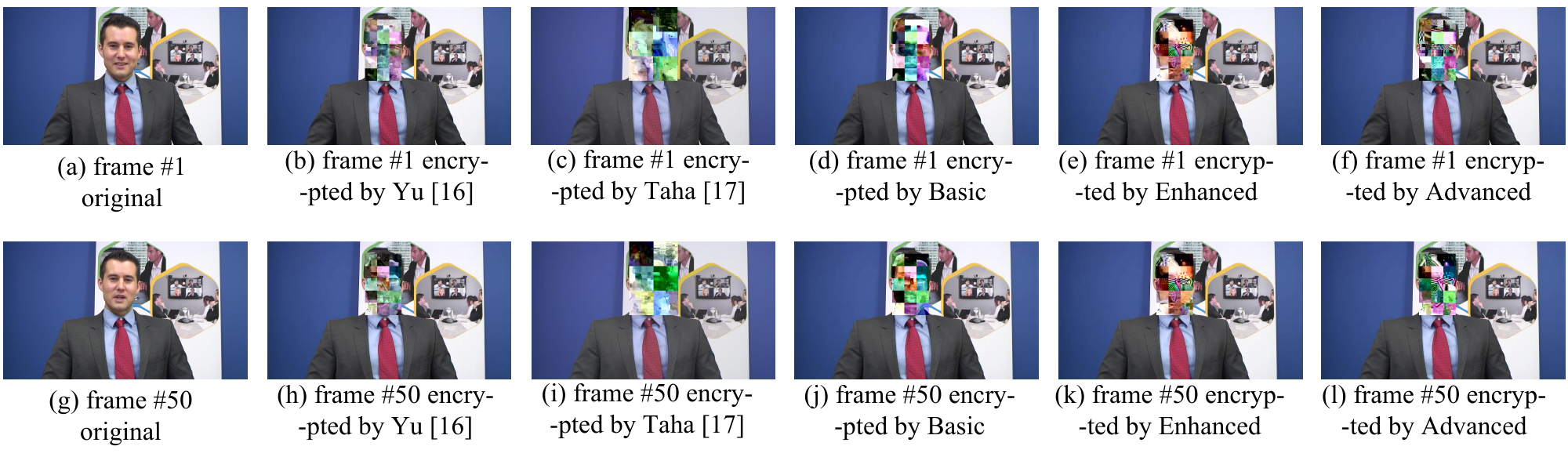}
    \caption{\small Subjective vision comparison of the whole frames in Johnny sequence}
    \label{fig7}
\end{figure}

\begin{figure}[htbp]
    \centering
    \includegraphics[width=0.7\linewidth]{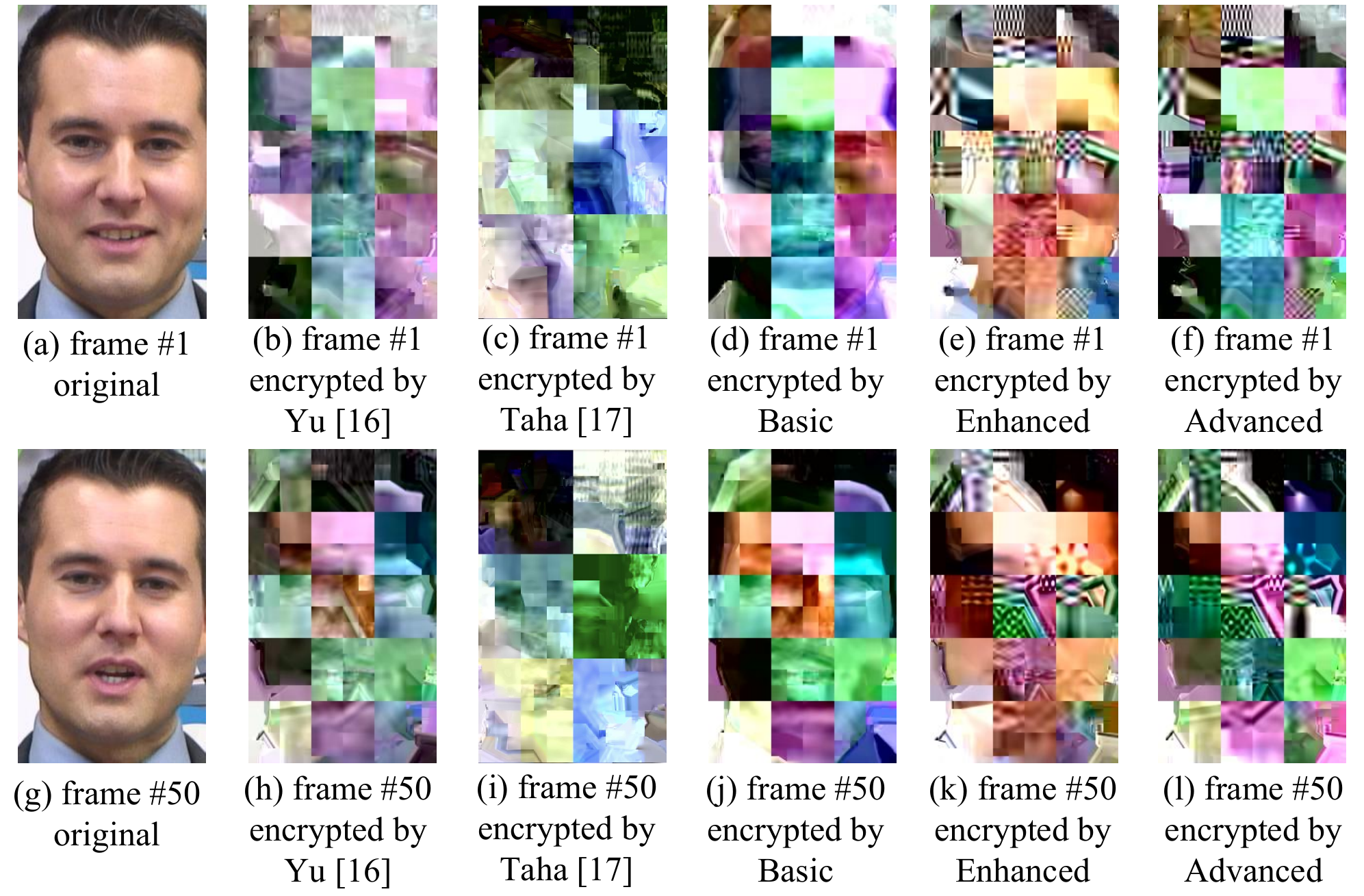}
    \caption{\small Subjective vision comparison of the ROI region in Johnny sequence}
    \label{fig8}
\end{figure}

\noindent\textbf{Comparison of Subjective Vision.}
According to \ref{SV}, the frame \#1 and \#50 of the Johnny sequence encrypted by five encryption strategies under $QP=24$ are listed in Fig. \ref{fig7}. Meanwhile, In order to present the distortion difference more clearly, we further slice the ROI region and make a detailed comparison of the distortion effect in Fig. \ref{fig8}. It can be clearly seen from Fig. \ref{fig8} that the visual distortion of Fig. \ref{fig8}(f) and Fig. \ref{fig8}(l) are the most serious, especially the texture details and contours of the face are almost unrecognizable. This is because the advanced level encryption uses edge scrambling in addition to the encryption syntax elements, causing greater distortion. However, since Yu\cite{yu2023coding} and Taha\cite{taha2018end} both encrypt the IPM, their encryption performance is better than that of the basic level encryption but worse than our enhanced level encryption. The reason is that enhanced level encryption introduces the encryption of Delta QP and MVPIdx on the basis of basic level encryption which can produce stronger visual distortion. Meanwhile, it can be seen from Fig.\ref{fig8} that the ROI region detected by the Taha\cite{taha2018end} is coarser than ours, containing more background areas, consistent with the results in Fig.\ref{fig1}.

\noindent\textbf{Comparison of Objective Indicators.}
According to TABLE \ref{tab3}, we evaluate the objective disturbance effects from the seven indicators: PSNR, SSIM, EDR, information entropy, NPCR, UACI and Bit rate change.
\paragraph{Analysis of PSNR and SSIM}
The testing results of PSNR and SSIM are recorded in TABLE \ref{tab:psnr} and TABLE \ref{tab:ssim}, where the optimal results are marked in bold and the suboptimal results are underlined. The results in the following tables of the other indicators are also highlighted in the same way. As can be seen from the tables, the PSNR and SSIM results of the advanced level encryption are almost the optimal and that of the enhanced level encryption are suboptimal. They are both better than that of Yu \cite{yu2023coding} and Taha \cite{taha2018end}.

\begin{table}[htbp]
\centering
\caption{\small Average PSNR of the compared strategies under different QP}
\resizebox{\linewidth}{!}{
\begin{tabular}{cccccccc}
\hline
\hline
Sequence & QP & Original & Yu\cite{yu2023coding} & Taha\cite{taha2018end} & Basic & Enhanced & Advanced \\
\hline
\multirow{3}{*}{CLASS\_A} & 8  & 51.47 & \underline{11.39} & \textbf{11.30} & 11.55 & 11.41 & 11.69 \\
                           & 24 & 40.56 & 11.49 & 11.60 & 11.58 & \underline{11.47} & \textbf{11.20} \\
                           & 40 & 31.62 & 11.69 & 11.82 & 12.04 & \textbf{11.43} & \underline{11.49} \\
\multirow{3}{*}{CLASS\_B} & 8  & 51.47 & 13.25 & 13.16 & 13.11 & \underline{13.05} & \textbf{12.86} \\
                           & 24 & 40.16 & \underline{13.50} & 13.60 & 13.55 & 13.53 & \textbf{12.49} \\
                           & 40 & 31.46 & 13.65 & 13.51 & 13.64 & \underline{13.40} & \textbf{12.85} \\
\multirow{3}{*}{CLASS\_C} & 8  & 51.67 & 13.14 & \underline{13.11} & 13.22 & 13.14 & \textbf{12.21} \\
                           & 24 & 44.97 & 13.42 & \underline{13.34} & 13.43 & 13.35 & \textbf{12.01} \\
                           & 40 & 38.10 & 12.97 & 13.12 & 13.09 & \underline{12.86} & \textbf{11.86} \\
\multirow{3}{*}{CLASS\_D} & 8  & 51.38 & 13.43 & 13.32 & 13.48 & \underline{13.19} & \textbf{12.72} \\
                           & 24 & 42.42 & 13.58 & \underline{13.50} & 13.53 & 13.63 & \textbf{13.26} \\
                           & 40 & 36.60 & 13.76 & \underline{13.61} & 13.67 & 13.65 & \textbf{13.54} \\
\multirow{3}{*}{CLASS\_E} & 8  & 51.38 & 12.44 & 12.55 & \underline{12.38} & 12.47 & \textbf{12.10} \\
                           & 24 & 41.93 & 12.76 & 12.66 & \underline{12.58} & 12.64 & \textbf{12.12} \\
                           & 40 & 37.51 & 12.95 & 13.06 & \underline{12.82} & 12.83 & \textbf{11.94} \\
\hline
\hline
\end{tabular}
\label{tab:psnr}
}
\end{table}

\begin{table}[htbp]
\centering
\caption{\small Average SSIM of the compared strategies under different QP}
\resizebox{\linewidth}{!}{
\begin{tabular}{cccccccc}
\hline
\hline
Sequence & QP & Original & Yu\cite{yu2023coding} & Taha\cite{taha2018end} & Basic & Enhanced & Advanced \\
\hline
\multirow{3}{*}{CLASS\_A} & 8  & 0.997 & 0.160 & 0.161 & 0.173 & \underline{0.137} & \textbf{0.131} \\
                           & 24 & 0.972 & 0.195 & 0.194 & 0.206 & \underline{0.178} & \textbf{0.155} \\
                           & 40 & 0.903 & 0.217 & 0.218 & 0.245 & \textbf{0.206} & \underline{0.213} \\
\multirow{3}{*}{CLASS\_B} & 8  & 0.998 & 0.180 & 0.178 & 0.189 & \underline{0.161} & \textbf{0.145} \\
                           & 24 & 0.996 & 0.220 & 0.221 & 0.230 & \underline{0.204} & \textbf{0.185} \\
                           & 40 & 0.820 & 0.256 & 0.253 & 0.263 & \underline{0.239} & \textbf{0.217} \\
\multirow{3}{*}{CLASS\_C} & 8  & 0.995 & 0.389 & 0.392 & 0.410 & \underline{0.364} & \textbf{0.323} \\
                           & 24 & 0.971 & 0.476 & 0.473 & 0.503 & \underline{0.458} & \textbf{0.410} \\
                           & 40 & 0.919 & 0.496 & 0.501 & 0.504 & \underline{0.475} & \textbf{0.402} \\
\multirow{3}{*}{CLASS\_D} & 8  & 0.996 & 0.435 & 0.431 & 0.428 & \underline{0.409} & \textbf{0.369} \\
                           & 24 & 0.943 & 0.472 & 0.477 & 0.457 & \underline{0.433} & \textbf{0.384} \\
                           & 40 & 0.876 & 0.510 & 0.497 & 0.510 & \underline{0.486} & \textbf{0.442} \\
\multirow{3}{*}{CLASS\_E} & 8  & 0.997 & 0.250 & 0.252 & 0.263 & \underline{0.231} & \textbf{0.204} \\
                           & 24 & 0.969 & 0.314 & 0.312 & 0.327 & \underline{0.289} & \textbf{0.259} \\
                           & 40 & 0.856 & 0.325 & 0.327 & 0.338 & \underline{0.306} & \textbf{0.274} \\
\hline
\hline
\end{tabular}
\label{tab:ssim}
}
\end{table}

\begin{figure}[htbp]
    \centering
    \includegraphics[width=0.8\linewidth]{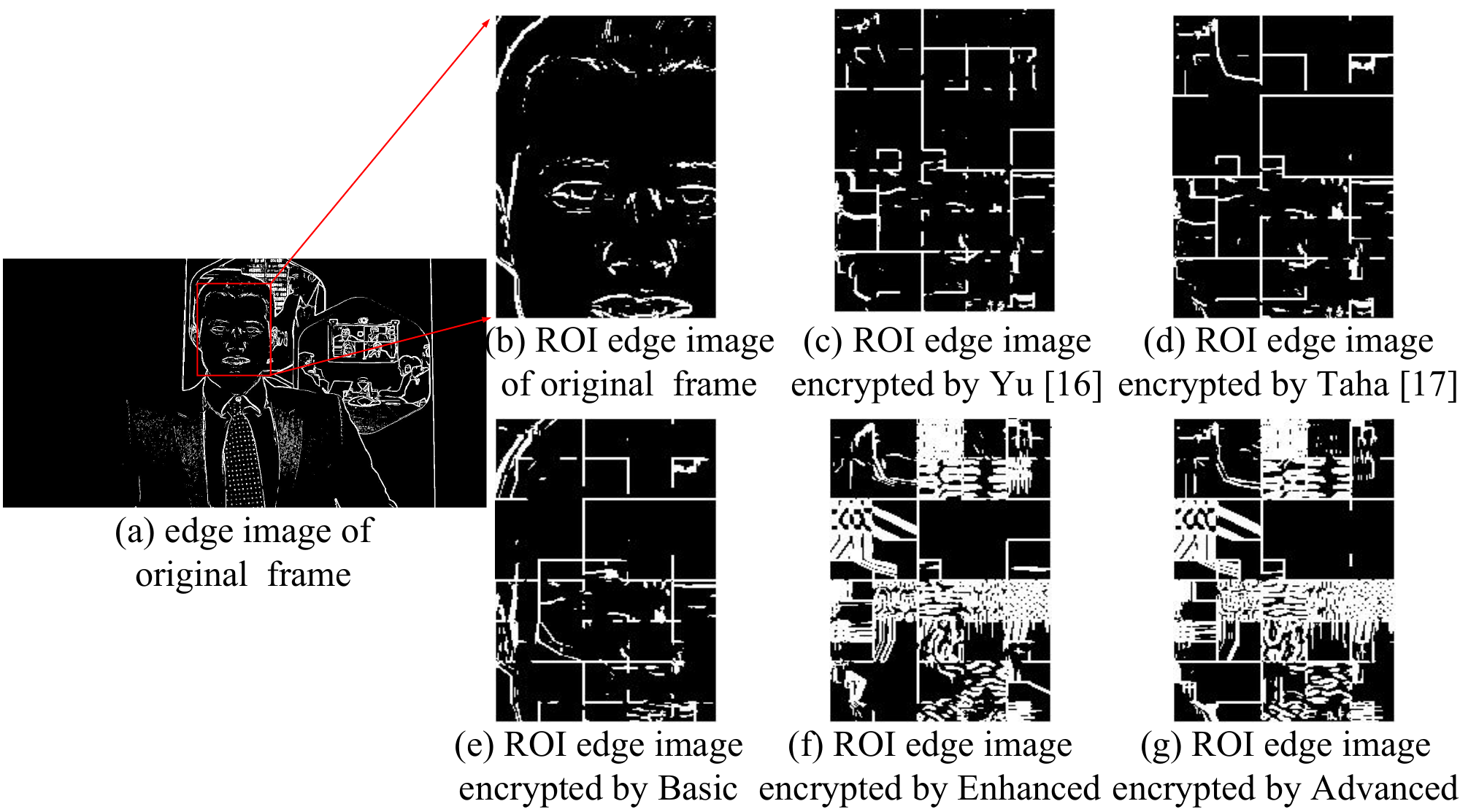}
     \caption{\small Edge image comparison of the ROI region in frame \#1 in Johnny sequence}
    \label{fig9}
\end{figure}

\paragraph{Analysis of Edge Detection}
we first give the encrypted visual effect in edge region of all the compared strategies in Fig. \ref{fig9}. It can be found that Yu\cite{yu2023coding} and Taha\cite{taha2018end} have large edge distortion than the basic level encryption resulting from the perturbation of IPM. However, both the edge distortion of enhanced and advanced level encryption are better than Yu\cite{yu2023coding} and Taha\cite{taha2018end}. The reason is that the enhanced level encryption encrypts MVPIdx and Delta QP and advanced level encryption scramble the edge coefficients, so the advanced level encryption reaches the best results. Meanwhile, the average values of EDR are also listed in TABLE \ref{tab6}. From the table, the advanced level encryption and the enhanced level encryption achieve the optimal and suboptimal results, which are better than Yu\cite{yu2023coding} and Taha\cite{taha2018end}. The result is consistent with the edge image in Fig \ref{fig9}.

\begin{table}[htbp]
\caption{\small Average EDR of the compared strategies}
\centering 
\resizebox{\columnwidth}{!}{%
\begin{tabular}{ccccccc}
\hline
\hline
\multirow{2}{*}{Sequence} & \multicolumn{6}{c}{EDR} \\ \cline{2-7} 
 & Original & Yu\cite{yu2023coding} & Taha\cite{taha2018end} & Basic & Enhanced & Advanced \\ \hline
CLASS\_A & 0.11 & 0.871 & 0.874 & 0.824 & \underline{0.919} & \textbf{0.926} \\
CLASS\_B & 0.132 & 0.849 & 0.849 & 0.789 & \underline{0.901} & \textbf{0.908}\\
CLASS\_C & 0.163 & 0.841 & 0.841 & 0.789 & \textbf{0.902} & \underline{0.901} \\
CLASS\_D & 0.307 & 0.905 & 0.907 & 0.831 & \underline{0.944} & \textbf{0.947} \\
CLASS\_E & 0.188 & 0.881 & 0.882 & 0.808 & \underline{0.922} & \textbf{0.923} \\ \hline
\hline
\end{tabular}
}
\label{tab6}
\end{table}

\begin{figure}[htbp]
    \centering
    \includegraphics[width=0.6\linewidth]{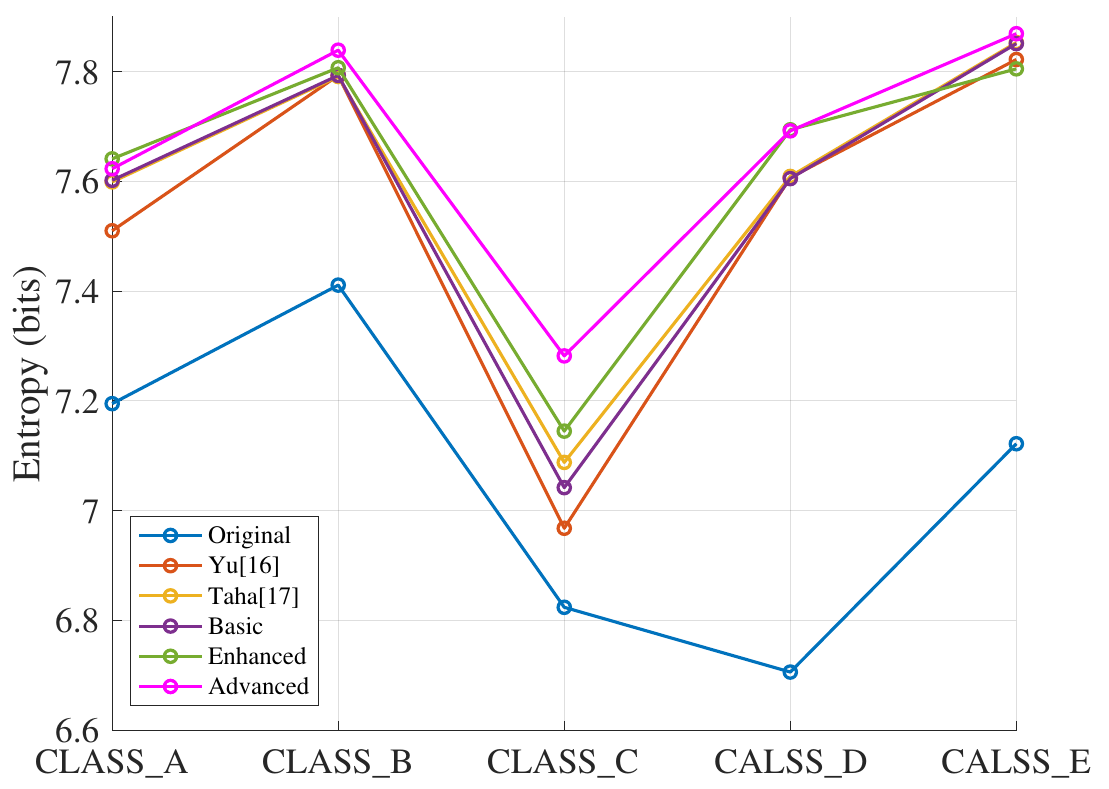}
    \caption{\small The comparison of information entropy}
    \label{fig10}
\end{figure}

\paragraph{Analysis of Information Entropy}
the results of the average information entropy of different video sequences in all the compared strategies are shown in Fig. \ref{fig10}. It can be seen that the information entropy of the original video is relatively low, ranging from 6.8 to 7.4, and the information entropy of Yu\cite{yu2023coding} and Taha\cite{taha2018end} is higher. However, the advanced level encryption achieve the highest information entropy, and the result of CLASS\_E reached approximately 7.95, close to the theoretical maximum value of 8.

\begin{table*}[htbp]
\centering
\caption{\small Average NPCR and UACI of the compared strategies}
\begin{tabular}{ccccccccccccc}
\hline
\hline
\multirow{2}{*}{Sequence} & \multicolumn{6}{c}{NPCR (\%)} & \multicolumn{6}{c}{UACI (\%)} \\ \cline{2-13} 
 & original & Yu~\cite{yu2023coding} & Taha~\cite{taha2018end} & Basic & Enhanced & Advanced & original & Yu~\cite{yu2023coding} & Taha~\cite{taha2018end} & Basic & Enhanced & Advanced \\
 \hline
CLASS\_A & 71.23 & \textbf{99.58} & 99.56 & 99.56 & 99.57 & \underline{99.57} & 1.26 & \textbf{29.17} & 29.00 & 28.99 & \underline{29.03} & 28.16 \\
CLASS\_B & 71.42 & 99.40 & 99.35 & 99.35 & \underline{99.41} & \textbf{99.48} & 1.12 & 23.97 & 24.08 & 24.03 & \underline{24.64} & \textbf{25.17} \\
CLASS\_C & 71.73 & 99.45 & 99.47 & \underline{99.48} & 99.47 & \textbf{99.49} & 1.29 & 24.01 & \underline{24.54} & 24.53 & 24.13 & \textbf{24.79} \\
CLASS\_D & 66.21 & \underline{99.50} & 99.46 & 99.47 & 99.49 & \textbf{99.66} & 0.63 & \underline{27.67} & 27.33 & 27.11 & 27.28 & \textbf{30.53} \\
CLASS\_E & 69.72 & 99.49 & 99.56 & \textbf{99.57} & 99.53 & \underline{99.55} & 0.80 & 26.13 & \textbf{27.60} & \underline{27.35} & 26.67 & 26.55 \\ \hline
\hline
\end{tabular}                                                                                                                  
\label{tab7}
\end{table*}

\paragraph{Analysis of NPCR and UACI}
The testing results of NPCR and UACI of all the compared strategies are listed in TABLE \ref{tab7}. the theoretical values of NPCR and UACI are 99.609\% and 33.464\%, respectively. Therefore. It can be seen from the table that advanced level encryption obtain the optimal NPCR and UACI, which proves it has good resistance to differential attacks.

\paragraph{Analysis of Bit Rate Change}
The experimental results of bit rate change are shown in TABLE \ref{tab8}. From the analysis in Section \ref{3B}, whether the encryption of the syntax element will cause an increase in bit rate depends on its entropy coding mode. Encrypting the syntax elements coded by regular mode will inevitably cause an increase in bit rate, Since Yu\cite{yu2023coding}, Taha\cite{taha2018end}, the enhanced and advanced level encryption all encrypting the syntax elements coded by regular mode, lead to bit rate increment. However, the basic level encryption only encrypts the syntax elements coded by bypass mode, which getting the best results with a zero bit rate increment.

\begin{table}[htbp]
\caption{\small Average bit rate change of the compared strategies}
\label{tab:erd_values}
\centering 
\begin{tabular}{lccccc} 
\hline
\hline
\multirow{2}{*}{Sequence} & \multicolumn{5}{c}{Bit Rate Change (\%)} \\ 
\cline{2-6} 
 & Yu~\cite{yu2023coding} & Taha~\cite{taha2018end} & Basic & Enhanced & Advanced \\ 
\hline
CLASS\_A & 1.56 & 1.56 & 0 & 2.53 & 8.11 \\
CLASS\_B & 1.95 & 2.02 & 0 & 2.77 & 7.94 \\
CLASS\_C & 2.23 & 2.30 & 0 & 3.47 & 10.61 \\
CLASS\_D & 0.79 & 0.80 & 0 & 2.41 & 8.8 \\
CLASS\_E & 1.58 & 1.60 & 0 & 2.71 & 9.83 \\ 
\hline
\hline
\end{tabular}
\label{tab8}
\end{table}
\section{Conclusions}\label{sec7}

This paper proposes a tunable scheme and evaluation benchmark for H.265/HEVC ROI encryption, aiming to provide a flexible and efficient privacy protection solution for video applications. By integrating a visual perception network for accurate ROI recognition and a three-level tunable encryption strategy, the scheme is developed to flexibly adapt to varying security and resource requirements. Experimental results demonstrate the scheme's effectiveness in privacy protection while maintaining real-time performance. Additionally, we present a unified ROI encryption evaluation benchmark, providing a standardized quantitative platform for video ROI encryption research. Future research will focus on optimizing the compression efficiency of encryption algorithms, incorporating more advanced object detection technologies, and further improve our evaluation benchmark.

\bibliographystyle{ieeetr}
\bibliography{BibTex}

\end{document}